\documentclass[preprint,showpacs,noshowkeys,floatfix,prd,aps,nofootinbib]{revtex4}
\usepackage{epsfig} 
\usepackage{longtable} 
\usepackage{hyperref}
\usepackage{bm}

\newcommand{\SM}{Standard Model}

\newcommand{\Hpm}{\mbox{$\mathrm{H}^{\pm}$}}
\newcommand{\Hp}{\mbox{$\mathrm{H}^+$}}
\newcommand{\Hm}{\mbox{$\mathrm{H}^-$}}

\newcommand{\mHpm}{\mbox{$m_{\mathrm{H}^{\pm}}$}}

\newcommand{\mW}{\mbox{$m_{\mathrm{W}^{\pm}}$}}
\newcommand{\mtop}{\mbox{$m_{\mathrm{t}}$}}

\newcommand{\mb}{\mbox{$m_{\mathrm{b}}$}}

\newcommand{\G}{\mbox{$\mathrm{GeV}$}}

\newcommand{\mA}         {\mbox{$m_{\mathrm{A}}$}}

\newcommand {\Ho}        {\mbox{$\mathrm{H}^{0}$}}
\newcommand {\Ao}        {\mbox{$\mathrm{A}^{0}$}}
\newcommand {\ho}        {\mbox{$\mathrm{h}^{0}$}}

\newcommand {\Wpm}       {\ensuremath{\mathrm{W}^{\pm}}}

\newcommand{\tanb}       {\mbox{$\tan\!\beta$}}

\newcommand{\gsim}{\;\raisebox{-0.9ex}
           {$\textstyle\stackrel{\textstyle >}{\sim}$}\;}

\newcommand{\lep}{\textsc{LEP}}
\newcommand{\cdf}{\textsc{CDF}}
\newcommand{\dzero}{\textsc{D0}}

\newcommand{\atlas}{\textsc{atlas}}
\newcommand{\atlfast}{\textsc{atlfast}}
\newcommand{\lhc}{\textsc{lhc}}
\newcommand{\herwig}{\textsc{herwig}}
\newcommand{\pythia}{\textsc{pythia}}
\newcommand{\acermc}{\textsc{AcerMC}}
\newcommand{\athena}{\textsc{athena}}
\newcommand{\ifb}{\ensuremath{\mathrm{fb}^{-1}}}
\newcommand {\Whad}  {\ensuremath{\mathrm{W}^{\pm}_{\mathrm{had}}}}
\newcommand {\Wlep}  {\ensuremath{\mathrm{W}^{\pm}_{\mathrm{lep}}}}
\newcommand {\thad}  {\ensuremath{\mathrm{t}_{\mathrm{had}}}}
\newcommand {\tlep}  {\ensuremath{\mathrm{t}_{\mathrm{lep}}}}
\newcommand {\deltaR}{\ensuremath{\Delta\mathrm{R}}}
\newcommand {\pT}    {\ensuremath{p_T}}
\newcommand {\bzero}    {\ensuremath{b_0}}

\newcommand {\btwo}    {\ensuremath{b_2}}

\newcommand {\tone}    {\ensuremath{t_1}}

\begin{document}  

\title{
The ATLAS discovery potential for a heavy charged Higgs boson in 
\bm{$gg \to tbH^\pm$} with \bm{$H^\pm \to tb$}}
\author{K\'et\'evi~A. Assamagan}  
\email{ketevi@bnl.gov} 
\affiliation{Department of Physics, Brookhaven National Laboratory, 
Upton, NY 11973 USA}
\author{Nils Gollub}
\email{nils.gollub@tsl.uu.se}
\affiliation{Department of Radiation Sciences,Uppsala University,
Box 535, 751~21 Uppsala, Sweden}
\date{February, 2004}  
\pacs{ 14.80.Cp, 12.60.Jv, 11.10.Kk} 
\keywords{charged Higgs; MSSM}

\def\OOrd{\lower .7ex\hbox{$\;\stackrel{\textstyle >}{\sim}\;$}}
\def\OOle{\lower .7ex\hbox{$\;\stackrel{\textstyle <}{\sim}\;$}}

\begin{abstract}    
The feasibility of detecting a heavy charged Higgs boson, 
\mbox{$\mHpm>\mtop+m_{b}$}, decaying in the \mbox{$\Hpm\rightarrow tb$} 
channel is studied with the fast simulation of the \atlas\
detector. 
We study the \mbox{$gg\rightarrow{\Hpm}tb$} production
process at the \lhc\ which together with the aforementioned decay
channel leads to four $b$--quarks in the final state. The whole
production and decay chain reads 
\mbox{$gg\rightarrow tb\Hpm\rightarrow	
t\bar{t}b\bar{b}\rightarrow	
b\bar{b}b\bar{b}l\nu\bar{q}q'$}.
Combinatorial background is a major difficulty in this multi--jet
environment but can be overcome by employing multivariate techniques
in the event reconstruction.
Requiring four $b$--tagged jets in the event helps to effectively suppress
the \SM\ backgrounds but leads to no significant improvement in the
discovery potential compared to
analyses requiring only three $b$--tagged jets.
This study indicates that charged Higgs bosons can be discovered at
the \lhc\ up to high masses 
(\mbox{$\mHpm>400\,\G$}) 
in the case of large \tanb.
\end{abstract}  
 
\maketitle  

\section{Introduction}
\label{sec:introduction}
The only particle predicted by the Standard Model (SM) 
that has so far not been detected
is the Higgs boson. 
In the Minimal Supersymmetric extension to the Standard Model
(MSSM)~\cite{Nilles:1984ge,Haber:1985rc}
the Higgs sector is enlarged to contain 5 particles: 3 neutral 
(\ho,\Ho,\Ao) and two charged (\Hp,\Hm) Higgs bosons. 
Whereas the detection of 
one neutral Higgs bosons 
would be compatible with both the SM and the MSSM, 
the detection of a charged
\mbox{spin--0} particle such as the charged Higgs boson predicted by the MSSM
would unequivocally point towards new physics beyond the SM.
This note describes the potential of the \atlas\ experiment to detect
a heavy charged Higgs boson, i.e. a charged Higgs boson heavier than 
the top quark, decaying in the \mbox{$\Hpm\rightarrow tb$} channel.

Other experiments have also searched for the charged Higgs boson and
set lower limits on the charged Higgs boson mass.  The combined \lep\
experiments provide a preliminary exclusion of charged Higgs bosons
with \mbox{$\mHpm<78.6\,\G$} at the
\mbox{$95\,\%$}~CL~\cite{unknown:2001xy}.  At the Tevatron, \cdf\ and
\dzero\ searched for the charged Higgs boson in the decay of top
quarks produced in \mbox{$p\bar{p}\rightarrow t\bar{t}$} reactions.
These searches exclude the low and high \tanb\ regions up to charged
Higgs masses of
\mbox{$\approx 160\,\G$}~\cite{Affolder:1999au,Abazov:2001md}.

\begin{figure}[]
  \begin{center}
    \epsfig{file=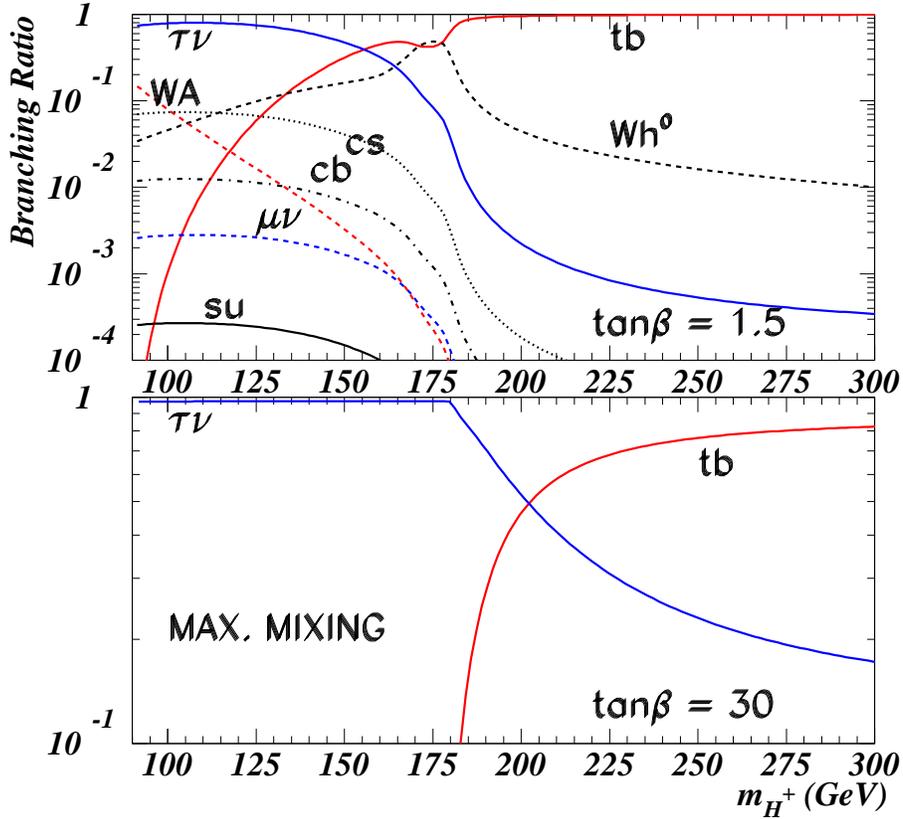,width=.8\textwidth}
  \end{center}
\vspace{-10mm}
\caption[]{\label{fig:BR}\sl 
Branching ratios for different decay channels of the charged Higgs
boson in the MSSM as a function of the charged Higgs mass. The upper
plot assumes $\tanb=1.5$, and the lower plot $\tanb=30$.
}  
\end{figure}

The Higgs sector of the MSSM is determined by two free parameters at
tree level, most often chosen to be the mass of the CP--odd neutral
Higgs boson, \mA, and the ratio of the vacuum expectation
values of the two electroweak Higgs doublets, parametrised by \tanb.
The decay modes of the charged Higgs boson in the MSSM 
are given as a function of 
the charged Higgs mass
in Figure~\ref{fig:BR} 
for two different values of $\tanb$, $1.5$ and $30$. 
These plots show that 
the main decay channel of heavy charged Higgs bosons for 
\mbox{$\mHpm\gsim\mtop+m_{b}$} 
is the decay into a top-- and a $b$--quark .
However, searches in this decay channel have to resolve the problem of 
a large multi--jet background.
For this reason the most promising channel for the search for  
the charged Higgs boson heavier than
the top quark is the \mbox{$\Hp\rightarrow\tau\nu_{\tau}$} 
decay channel, as it provides a lower background 
environment~\cite{Assamagan:2002in}.

The \mbox{$\Hpm\rightarrow tb$} decay channel, assuming
\mbox{$\mHpm\gsim\mtop+m_{b}$}, has been studied in a previous
note~\cite{Assamagan:2000uc}, using the \mbox{$2\rightarrow 2$}
production process \mbox{$gb\rightarrow \Hpm t$} and detecting 3
$b$--jets in the final state. However, as was shown in that report, the
large background from \SM\ $t\bar{t}$--production complicates the
detection of the charged Higgs boson and limits the mass region for a
charged Higgs discovery to masses below about \mbox{$400\,\G$} for low
or high values of \tanb.  The purpose of the present study is to try
to extend the discovery reach beyond this limit.

Recently it was suggested~\cite{Miller:1999bm} that by utilizing the
$2\rightarrow 3$ production process $gg\rightarrow \Hpm tb$ in
combination with the \mbox{$\Hpm\rightarrow tb$} decay, the fourth $b$--jet
inherent in the signal process could be detected,  
resulting in a greater rejection 
of the \SM\ background processes.
We therefore study a heavy charged Higgs boson in the
production and decay chain
\begin{equation}
\label{eq:signal}
gg\rightarrow tb\Hpm\rightarrow 
t\bar{t}b\bar{b}\rightarrow 
b\bar{b}b\bar{b}l\nu\bar{q}q',
\end{equation}
where one of the top quarks is required to decay leptonically in order
to provide a hard isolated lepton to trigger on.
The SM background processes
that lead to the same final state with four $b$--tagged jets are
\begin{equation}
gg\rightarrow t\bar{t}b\bar{b}
\end{equation}
and
\begin{equation}
gg\rightarrow t\bar{t}gg + t\bar{t}q\bar{q},
\end{equation}
where, in the latter case, two of the light jets are misidentified as
$b$--jets.

\begin{figure}[]
  \begin{center}
    \epsfig{file=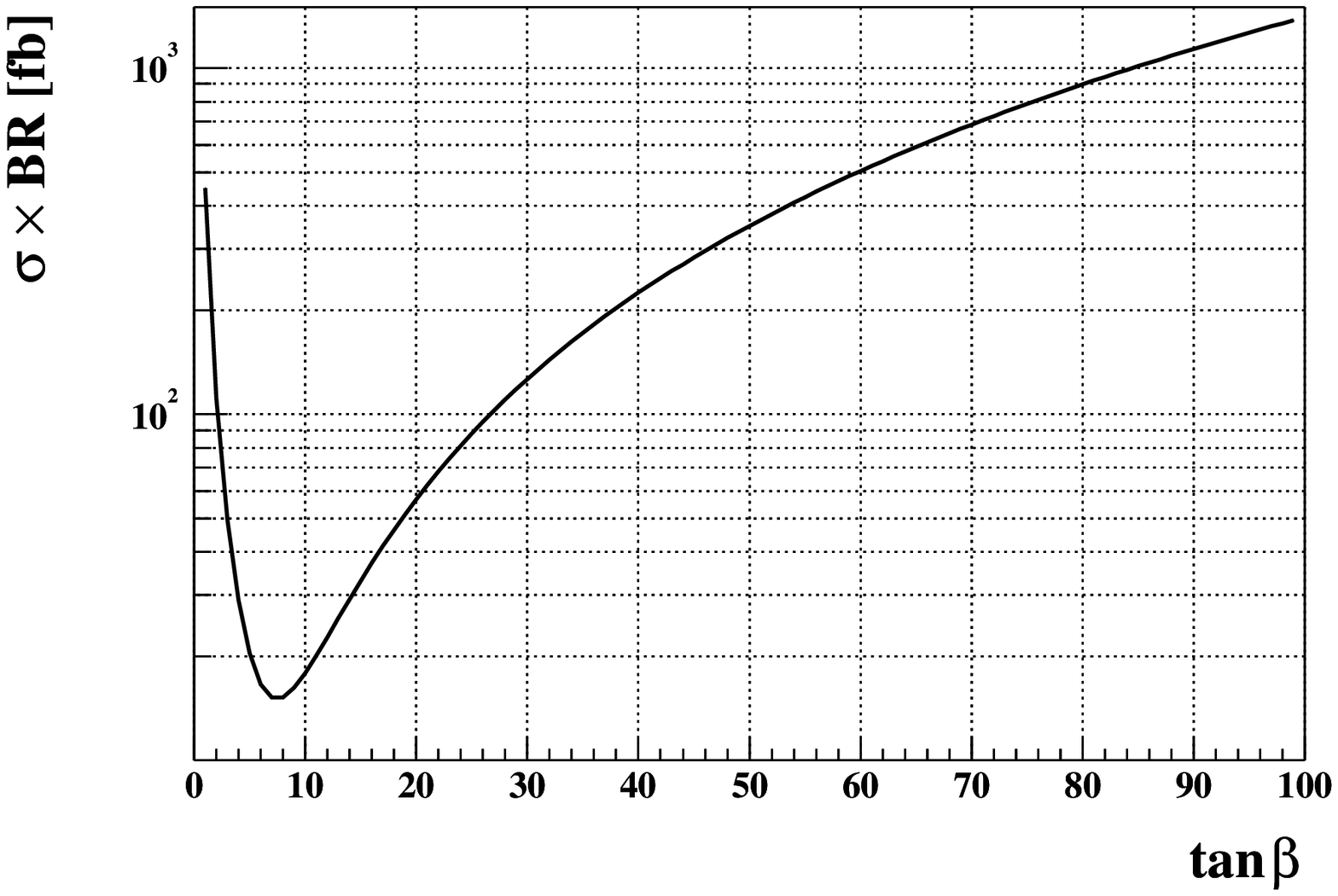,width=.7\textwidth}
  \end{center}
\vspace{-5mm}
\caption[]{\label{fig:sigma_tanb}\sl 
\mbox{$\sigma\times\mathrm{BR}$} of the process 
\mbox{$gg\rightarrow tb\Hpm\rightarrow 
t\bar{t}b\bar{b}\rightarrow 
b\bar{b}b\bar{b}l\nu\bar{q}q'$} 
for a
charged Higgs mass of \mbox{$\mHpm=300\,\G$} as a function of \tanb.  
}  
\end{figure}

Analyses searching for a charged Higgs boson in the 
production processes 
\mbox{$gb\rightarrow{\Hpm}t$} and
\mbox{$gg\rightarrow{\Hpm}tb$}
generally suffer from a lack of
sensitivity for intermediate values of \tanb.
The relevant part of the MSSM Lagrangian describing the
${\Hpm}tb$ Yukawa coupling is given by~\cite{Miller:1999bm}
\begin{equation}
{\cal L}=\frac{e}{\sqrt{2}\mW\sin\theta_{W}}\Hp
\left(
\mb\tanb\,\bar{t}\,b_{R}+\mtop\cot\!\beta\,\bar{t}\,b_{L}
\right),
\end{equation}
which has a minimum at \mbox{$\tanb=\sqrt{\mtop/\mb}$}. 
This behavior is illustrated in 
Figure~\ref{fig:sigma_tanb} 
showing the cross section times branching ratio (BR) 
for process~(\ref{eq:signal}) 
as a function of \tanb.
The dip around \mbox{$\tanb=\sqrt{\mtop/\mb}\approx7$} is apparent.

In the following section the event generation and detector simulation
are described. Section~\ref{sec:analysis} describes the analysis which
is divided into two likelihood selections presented in 
the sections~\ref{sec:combLik} and~\ref{sec:selLik} respectively.
Section~\ref{sec:results} summarises the results, and conclusions and
an outlook are given in section~\ref{sec:conclusion}.

\section{Event Generation and Simulation}
\label{sec:simulation}
The event generation and detector simulation for the various Monte
Carlo (MC) samples
used in this analysis are done within the \athena\ framework
in the \atlas\ Software~Release~6.0.3.

The signal process (\ref{eq:signal}) 
is generated for a charged Higgs boson mass range of
\mbox{$\mHpm=200-800\,\G$}
using \herwig~6.5~\cite{HERWIG,HERWIGSUSY}. 
Table~\ref{tab:hProd} lists the mass points at which MC
samples are produced and gives production cross sections for a set
of \tanb\ values.
In the cross section calculations the factorisation and
renormalisation scales, $\mu_F$ and $\mu_R$, respectively,
are set to the mean transverse mass so that
$\mu_F^2=\mu_R^2
=\sum_{i=b,t} ({p_T^i}^2 + m_i^2)/2
=(m_{T}^2(t)+m_T^2(b))/2 =\left< m_T^2\right>$. 
As all cross sections are calculated at leading order and 
no next to leading order calculations exist to date, 
an optimal choice of the QCD scale is not obvious.
The choice of this particular scale is guided by the demand to use 
comparable scales in the signal and background calculations for
consistency
and by results obtained in \cite{gg_ttH} although the
process considered there can not be compared directly to the signal
process considered in this report. 
Choosing the QCD scale is one of the main systematic uncertainties
when predicting the signal and background cross sections. 
The value assumed here provides estimates of the cross sections to
be expected but can by no means be considered as definitive.
Other choices of the QCD scale or $m_b$ evaluation may result in
cross sections differing by up to a factor~2.
This topic is further discussed in section \ref{sec:results}.  
\begin{table*}
\begin{center}
\begin{minipage}{1\linewidth} 
\caption{\label{tab:hProd} 
\sl Summary of signal MC samples for the various charged Higgs masses
studied in this analysis. Inclusive and exclusive cross sections are 
quoted for a selection of \tanb\ values, assuming
\mbox{$\mu_F^2=\mu_R^2
=(m_{T}^2(t)+m_T^2(b))/2$} and a running $b$--quark mass.
Each MC sample contains $10^6$~generated events. 
}
\end{minipage}
\begin{tabular}{rrrrr} \hline
\mHpm\ & \ \ \ \ \tanb\  & \ \ \ \ $\sigma_{incl.}\,[pb]$ & \ \ \ \  
$\sigma\times\mathrm{BR}\,[pb]$ 
\\
\hline
200 & 30 & 1.2 & 0.17  \\
250 & 40 & 1.3 & 0.31  \\
300 & 50 & 1.4 & 0.35  \\
350 & 60 & 1.4 & 0.35  \\
400 & 20 & 0.11 & 0.029  \\
500 & 30 & 0.13 & 0.034  \\
600 & 40 & 0.13 & 0.033  \\
800 & 30 & 0.024 & 0.0063  \\
\hline 
\end{tabular}
\end{center}
\end{table*}

The strong coupling constant $\alpha_s$ is evaluated at the 1--loop level 
and a running $b$--quark mass is used.
A central value of \mbox{$m_t=175\,\G$} is assumed for the
top quark mass and the \textsc{cteq5l} parton density function is 
used throughout the analysis. 
\mbox{$\mathrm{H}^{+}\rightarrow t\bar{b}$} branching ratios  
are evaluated with HDECAY~3.0~\cite{Djouadi:1998yw} 
where the decay
to supersymmetric particles is switched off (OFF-SUSY=1).
We evaluate the branching ratios in the Maximal Mixing Scenario as 
described in~\cite{Carena:1999xa}, 
assuming the top quark mass mentioned above:
\mbox{$M_{\mathrm{SUSY}}=1\,\mathrm{TeV}$}, 
\mbox{$M_2=200\,\G$}, 
\mbox{$\mu=-200\,\G$}, 
\mbox{$M_{\tilde{g}}=800\,\G$} and 
\mbox{$A_t=A_b=X_t+\mu/\tanb$} with 
\mbox{$X_t=\sqrt{6}M_{\mathrm{SUSY}}$}.
The branching ratio of the \Wpm\ decaying to quarks is assumed to be
$2/3$, and the BR to a lepton (electron or muon) 
plus the accompanying neutrino
is taken to be $2/9$. 
Since each of the \Wpm\ can decay leptonically or hadronically, 
an overall factor of $2$ has to be applied. 
This leads to the following relation:
\mbox{$\sigma\times\mathrm{BR} = 
\sigma_{incl.}\times\mathrm{BR}(\mathrm{H}^{+}\rightarrow t\bar{b})
\times 8/27.$}

Samples for the background processes 
\mbox{$gg\rightarrow t\bar{t}b\bar{b}$},  
\mbox{$qq\rightarrow t\bar{t}b\bar{b}$} and
\mbox{$gg\rightarrow Z/\gamma/W\rightarrow t\bar{t}b\bar{b}$}
are produced with \acermc~1.2 \cite{Kersevan:2002dd} in stand--alone mode.
The QCD energy scale is chosen such that
\mbox{$Q^{2}_{QCD}=\Sigma ({p_T^i}^2 + m_i^2)/4=\left<m^{2}_{T}\right>$}.
These samples are then passed to \herwig~6.5 within the Athena
framework for fragmentation and hadronisation. 
In order to study the systematic uncertainty due to different
fragmentation schemes, the \mbox{$gg\rightarrow t\bar{t}b\bar{b}$}
sample is also
passed to \pythia~6.203~\cite{Sjostrand:2000wi} for a similar treatment.
The effect of mis--tagging light jets as $b$--jets  
is studied with the help of
a large \mbox{$t\bar{t}+\mathrm{jets}$} sample generated with \herwig~6.5.
Table~\ref{tab:bgdProd} summarises the background samples and
their inclusive cross sections.
\begin{table*}
\begin{center}
\begin{minipage}{1\linewidth} 
\caption{\label{tab:bgdProd} 
\sl Summary of background MC samples studied in this analysis. 
\vspace{4mm}
}
\end{minipage}
\begin{tabular}{llrr} \hline
Process  & Generator    & $\sigma_{incl.}\,[pb]$ & \ \ generated events \\
\hline
$gg\rightarrow t\bar{t}b\bar{b}$ & \acermc~1.2 & 10.3 & $10$\,M \\   
$qq\rightarrow t\bar{t}b\bar{b}$ & \acermc~1.2 &  0.61 & $1$\,M  \\
$gg\rightarrow Z/\gamma/W\rightarrow t\bar{t}b\bar{b}$ \ \ \ \ \ &
\acermc~1.2 & 1.1 & $1$\,M \\
$t\bar{t}+\mathrm{jets}$ & \herwig~6.5\ \ \ \  & 405.0 & $50$\,M\\
\hline 
\end{tabular}
\end{center}
\end{table*}

The \atlas\ detector is simulated with the fast detector simulation 
\atlfast\ as it is represented in the \atlas\ Software~Release~6.0.3. 
This package
is based on the \textsc{fortran} implementation of the same package
\cite{ATLFAST}.
Jets are reconstructed with a cone based algorithm using a cone size of
\mbox{$\deltaR=0.4$}. 
Only jets having a minimum transverse momentum of \mbox{$\pT>10\,\G$}
and lying in the pseudorapidity range of \mbox{$|\eta|<5$} are 
accepted for this analysis. 
An efficiency of \mbox{$90\,\%$} to identify isolated charged leptons is
assumed. Jet energy and momentum calibration and the $b$--tagging of
jets is performed within the \textsc{atlfastb} routine of the
\atlfast\ simulation package. 
The possibility to tag a jet as a $b$--jet is limited by the inner tracker
acceptance range of \mbox{$|\eta|<2.5$}.
A $b$--tagging efficiency of \mbox{$60\,\%$} is
assumed when simulating samples for the low luminosity option of the
\lhc, \mbox{$50\,\%$} for the high luminosity option.
Rejection factors of \mbox{$\mathrm{R}_\mathrm{c}=10$} and
\mbox{$\mathrm{R}_\mathrm{j}=100$} are chosen for c-- and light jets respectively. 
The $b$--tagging efficiencies and rejection factors are static,
i.\@e.\@ they do not depend on the pseudorapidity $\eta$ or 
transverse momentum $\pT$ of the jets.
All plots and tables shown in this analysis refer to the
low luminosity option of the \lhc, assuming an integrated luminosity
of \mbox{$30\,\ifb$} unless explicitly stated otherwise.

\subsection{Jet--Parton Matching}
\label{sec:jetParton}
In order to construct and test the performance of the event 
reconstruction algorithm
(see section \ref{sec:combLik}), it is necessary to know the link
between a generated parton and a detected jet or lepton. The former is
often referred to as the ``Monte Carlo truth'', and the latter will be
referred to as ``reconstructed objects'' in the following. 
Initially no such link between a parton and a reconstructed object
is provided by the MC generator or the detector simulation program 
and the association is far from straightforward. 
In this analysis the 
problem is handled approximately by solving the assignment problem 
as described in \cite{assignmentProblem}.
The quantity which is minimised is the sum of all distances between 
the generated partons after final state radiation (FSR) 
and their associated reconstructed object 4--vectors.
The distance between the 4--vector of a parton
and the 4--vector of a reconstructed object is given by 
\mbox{$\deltaR = \sqrt{\Delta^{2}\eta + \Delta^{2} \phi}$}, 
the distance in pseudorapidity--azimuthal angle space.
If the distance between a parton and its reconstructed object 
exceeds $0.4$ it is assumed that no association is possible.
Further, no association is attempted if any of the initial quarks
after FSR has a 4--momentum outside the acceptance range of the 
JetMaker algorithm in \atlfast. 

\section{Analysis}
\label{sec:analysis}
The analysis has three parts.
In the first step all events are required to pass a set of cuts 
in order to reject most of the SM background and to
ensure the minimum prerequisites needed for subsequent reconstruction.

The second part 
is intended to find the combination of jets that correctly reconstructs
the two top quarks and the
charged Higgs present in the final state of the signal process. For
each event the most likely correct combination is found with the help
of a selection procedure described in section \ref{sec:combLik}. This
likelihood is referred to as the ``combinatorial
likelihood''.

Once the correct combination is found for each event, a second
likelihood selection, the ``selection likelihood'' described in
section \ref{sec:selLik}, aims at separating the signal from the SM 
background processes.

\subsection{Preselection}
\label{sec:precuts}
In the preselection, events with a topology clearly distinct from the 
signal topology are rejected. 
This ensures that only the main backgrounds discussed in section
\ref{sec:introduction} need to be studied further. 
The preselection requires:
\begin{itemize}
\item
exactly 1 isolated lepton (\mbox{$l=\mathrm{e\ or}\ \mu$}) with 
transverse momentum \mbox{$\pT^{e}>25\,\G$}, \mbox{$\pT^{\mu}>20\,\G$} and
pseudorapidity \mbox{$|\eta|<2.5$},  
\item
exactly 4 $b$--jets with $|\eta|<2.5$ and \mbox{$\pT>20\,\G$} and 
\item 
at least 2 light jets with $|\eta|<5$ and \mbox{$\pT>20\,\G$}.
\end{itemize}
In order to trigger on the signal events the detection of a high--\pT\
lepton is required.
The cuts applied to the $\pT$ and $\eta$ of the isolated lepton 
are chosen such that they meet the requirements of the \atlas\ trigger
system. 
When running in the high luminosity option the cut on the jets'
transverse momenta is increased from \mbox{$\pT^{\mathrm{jet}}>20\,\G$} to 
\mbox{$\pT^{\mathrm{jet}}>30\,\G$}.
The efficiency of the precuts for signal events depends on the
assumed charged Higgs mass and ranges 
from \mbox{$1.78\,\%$} for \mbox{$\mHpm=200\,\G$}
to \mbox{$4.41\,\%$} for \mbox{$\mHpm=800\,\G$}. 
The precut efficiencies are summarised in Table~\ref{tab:preCutEff}.
\begin{table*}
\begin{center}
\begin{minipage}{1\linewidth} 
\caption{\label{tab:preCutEff} 
\sl 
Preselection efficiencies for the different charged Higgs masses studied 
in this analysis.
The efficiencies are quoted for the low luminosity (LL) and the 
high luminosity (HL) option of the \lhc.
\vspace{2mm}
}
\end{minipage}
\begin{tabular}{ccc} \hline
\ \ \ \ \mbox{$\mHpm$} [GeV]\ \ \ \  &
\ \ \ \  preselection efficiency, LL [\%]\ \ \ \ &
\ \ \ \  preselection efficiency, HL [\%] \\
\hline
200 & $1.783 \pm  0.013$ & $0.5687 \pm 0.0075$  \\ 
250 & $3.171 \pm  0.018$ & $1.121  \pm 0.011$   \\
300 & $3.577 \pm  0.019$ & $1.281  \pm 0.011$   \\
350 & $3.738 \pm  0.019$ & $1.436  \pm 0.012$   \\
400 & $3.932 \pm  0.019$ & $1.509  \pm 0.012$   \\
500 & $4.123 \pm  0.020$ & $1.644  \pm 0.013$   \\
600 & $4.281 \pm  0.020$ & $1.715  \pm 0.013$   \\
700 & $4.363 \pm  0.020$ &  (not studied)      \\
800 & $4.411 \pm  0.021$ & $1.872  \pm 0.014$   \\
\hline 
\end{tabular}
\end{center}
\end{table*}

In order to reconstruct the leptonically decaying \Wpm\ (\Wlep)
the 4--momentum of the daughter neutrino 
needs to be reconstructed. 
The x-- and y--components of the neutrino momentum are assumed to
coincide with the measured missing transverse momentum components 
$p_x^{\mathrm{miss}}$ and $p_y^{\mathrm{miss}}$ respectively.
The z--component however can not be measured but must be calculated
by solving the equation:
\[
m_{\Wpm}^{2}=(E_{\nu}+E_{l})^{2}-(\vec{p}_{\nu}+\vec{p}_{l})^{2} ,
\mathrm{\ \ \ with\ \  } E_{\nu}=|\vec{p_{\nu}}|.
\]
This equation can result in two or zero real solutions for $p_{\nu}^{z}$.
If two solutions are found, both are kept for later evaluation in
the event reconstruction algorithm.
However, in approximately \mbox{$25\,\%$} of the events 
no solution is found. 
In order to
keep these events and still be able to reconstruct the leptonically 
decaying \Wlep\ in those otherwise fatal cases, the collinear
approximation approach described in \cite{jochen} is adopted:
\mbox{$p_{\nu}^{z}=p_{l}^{z}$} is assumed if no solution can be found,
and the resulting \Wpm\ 4--momentum 
is rescaled to match $m_{\Wpm}$.
This increases the \Wlep\ reconstruction efficiency from \mbox{$75\,\%$} to 
\mbox{$100\,\%$} and only a small loss in the resolution of the reconstructed
leptonically decaying \tlep\ is observed.

\subsection{The Combinatorial Likelihood}
\label{sec:combLik}
The final state of the signal process (\ref{eq:signal})
is quite complex, featuring four $b$--jets, two light jets
from the hadronically decaying \Wpm (\Whad) 
and an isolated lepton plus missing transverse momentum
from the leptonically decaying \Wpm (\Wlep). 
Quark and gluon jets from initial and final state
radiation and the underlying event are also present,
increasing the jet multiplicity. 
Initially it is unknown which reconstructed objects should be 
combined to reconstruct the two {\Wpm}s, 
the two top quarks, and finally the charged Higgs boson. 
The combinatorial likelihood aims at identifying the correct reconstructed 
objects to combine and thereby making the correct reconstruction 
of the whole event possible.
In order to incorporate as much information available from each event
as possible a multivariate technique is chosen to find the correct
combination of reconstructed objects for each event.
We choose to implement a likelihood selection distinguishing
two classes where the first class represents the correct combination
and the second class all the wrong ones. 
The likelihood formalism used in this analysis is outlined briefly in the 
following, generalising to $n$ classes of events:

For each of the $m$ observables $x_i$ used to 
distinguish between the $n$ classes, 
the normalised probability density functions (pdf) 
\begin{equation}
  f_i^j(x_i),\qquad\text{ where } i=1,\dots,m\text{ and } j=1,\dots,n 
\end{equation}
have to be determined.
The probability that an event belongs to class $j$ when 
measuring the value $x_i$ for variable $i$ 
is given by
\begin{equation}
  p_i^j(x_i)=\frac{f_i^j(x_i)}{\sum^n_{k=1}f_i^k(x_i)}. \label{eq:likProb}
\end{equation}
The likelihood \mbox{${\cal L}$} that an event belongs to class j when measuring
$m$ variables $x_1,\dots,x_m$ is then given by
the normalised product of the probabilities Eq.\,\ref{eq:likProb} over all $m$ variables:  
\begin{equation}
  {\cal L}^j=
  \frac{\displaystyle \prod^m_{l=1}p_l^j(x_l)}{\displaystyle \sum^n_{k=1}\prod^m_{l=1} p_l^k(x_l)}.
\end{equation}
The likelihood has values in the range 0,\dots,1.
Information about possible correlations between the input variables is
neglected by this procedure.

When constructing the pdfs for the combinatorial likelihood 
it is essential to know the
correct association of partons to reconstructed objects as described 
in section~\ref{sec:jetParton}.
The fraction of signal events passing the precuts for which 
a valid association to partons is found depends
on the charged Higgs mass and
ranges from \mbox{$45\,\%$} at \mbox{$\mHpm=200\,\G$} 
to \mbox{$58\,\%$} at \mbox{$\mHpm=800\,\G$}.
Only events having a valid parton association can be used to obtain the 
pdfs for the correct-- and the wrong--combination class.

Any algorithm used to reconstruct the events  
has a chance to find the correct 
combination only if the correct four jets are $b$--tagged 
and the two light jets
originating from the hadronically decaying \Wpm\ pass the precut constraints.
Of those events passing the precuts and 
obtaining a valid association to partons, 
only approximately \mbox{$65\,\%$} fulfill this requirement. 
This means that for~\mbox{$\approx 35\,\%$} of signal events passing the precuts
the completely correct reconstruction is doomed from the beginning.

\begin{figure}[]
    \begin{minipage}[t]{1\textwidth}
      \begin{minipage}[t]{.32\textwidth}
        \begin{center}
          \epsfig{file=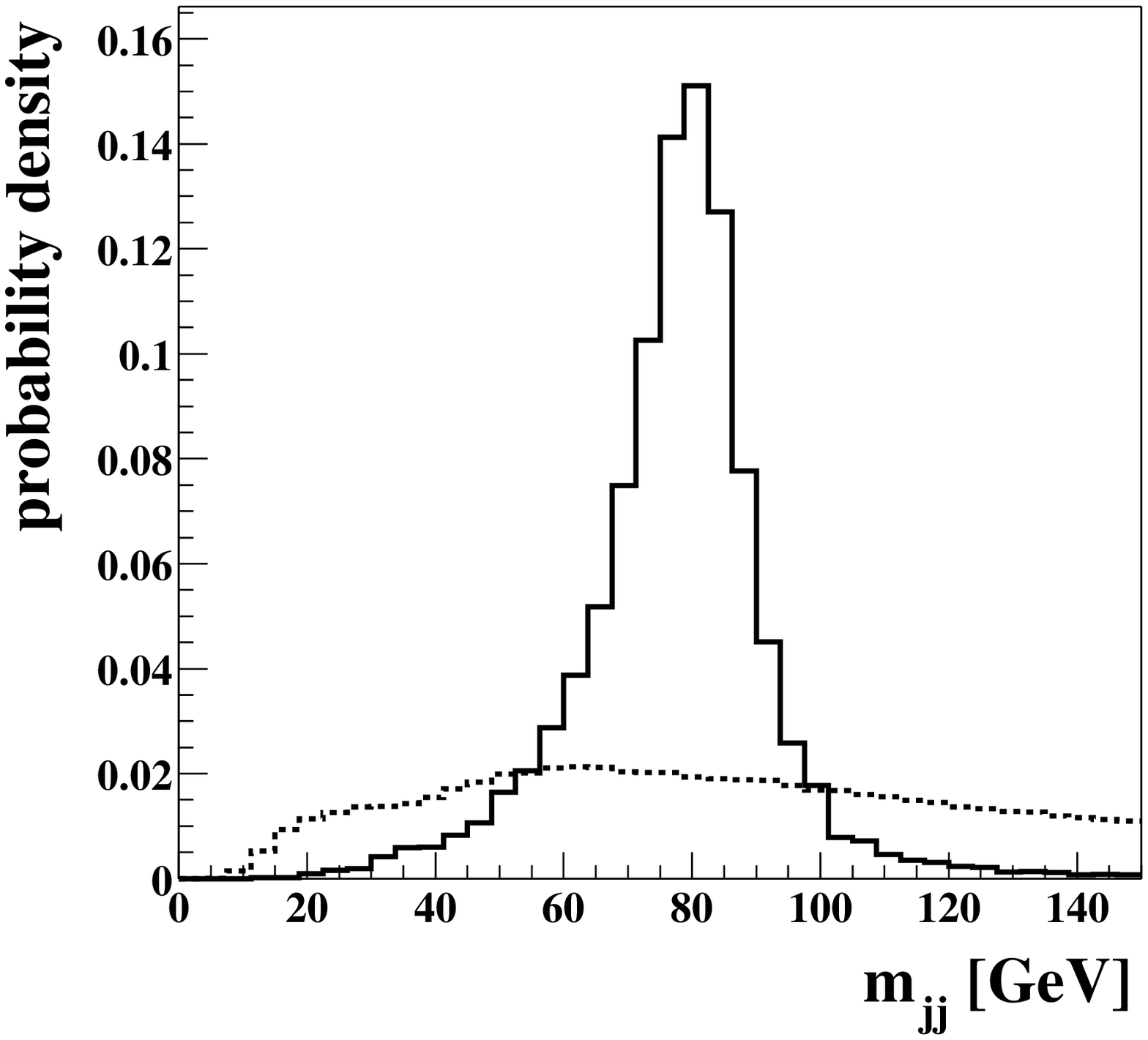,width=1.06\textwidth}
        \end{center}
      \end{minipage}
      \begin{minipage}[t]{.32\textwidth}
        \begin{center}
          \epsfig{file=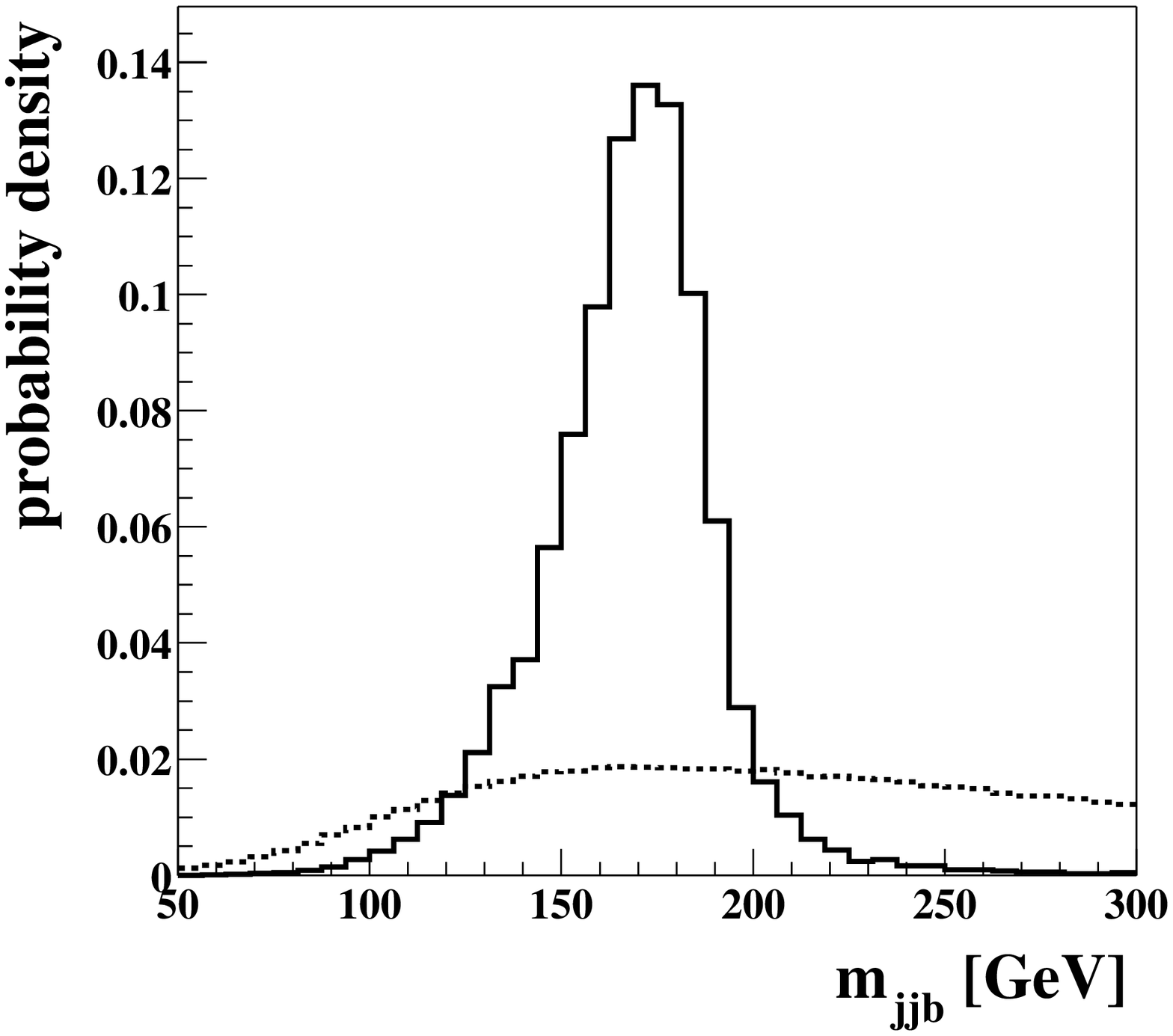,width=1.06\textwidth}
        \end{center}
      \end{minipage}
      \begin{minipage}[t]{.32\textwidth}
       \begin{center}
          \epsfig{file=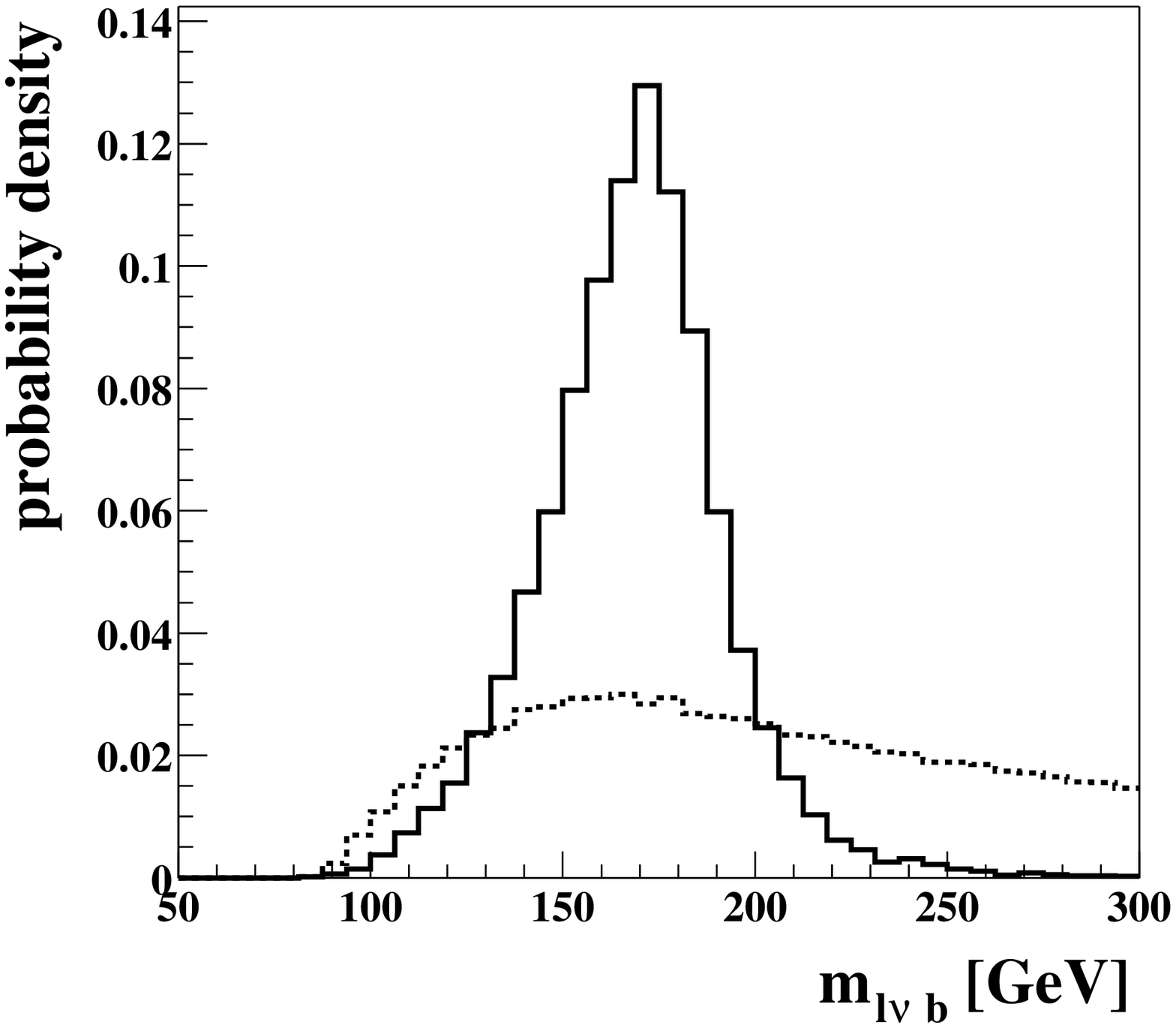,width=1.06\textwidth}
        \end{center}
      \end{minipage}
    \end{minipage}

    \begin{minipage}[t]{1\textwidth}
      \begin{minipage}[t]{.32\textwidth}
        \begin{center}
          \epsfig{file=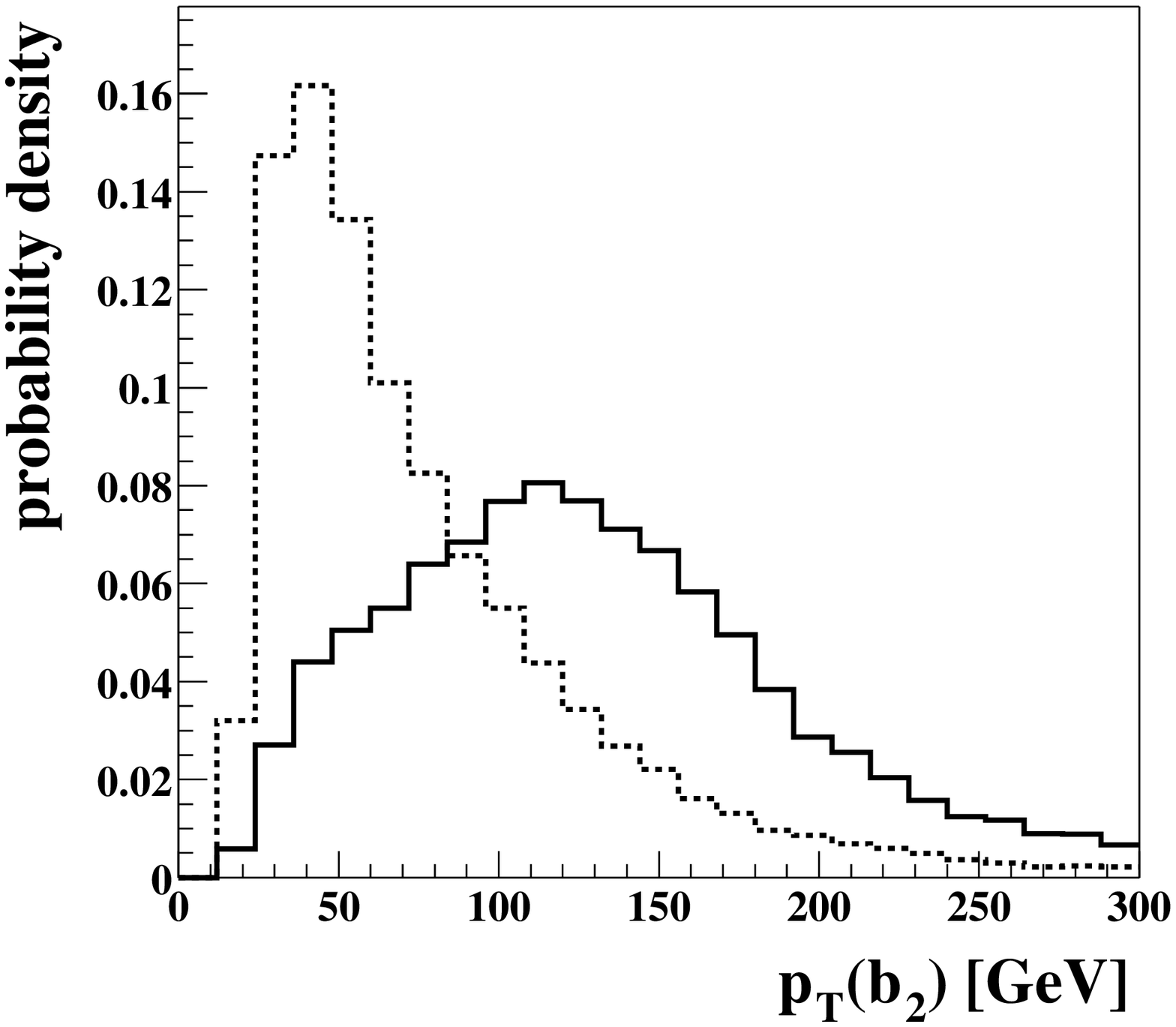,width=1.06\textwidth}
        \end{center}
      \end{minipage}
      \begin{minipage}[t]{.32\textwidth}
        \begin{center}
          \epsfig{file=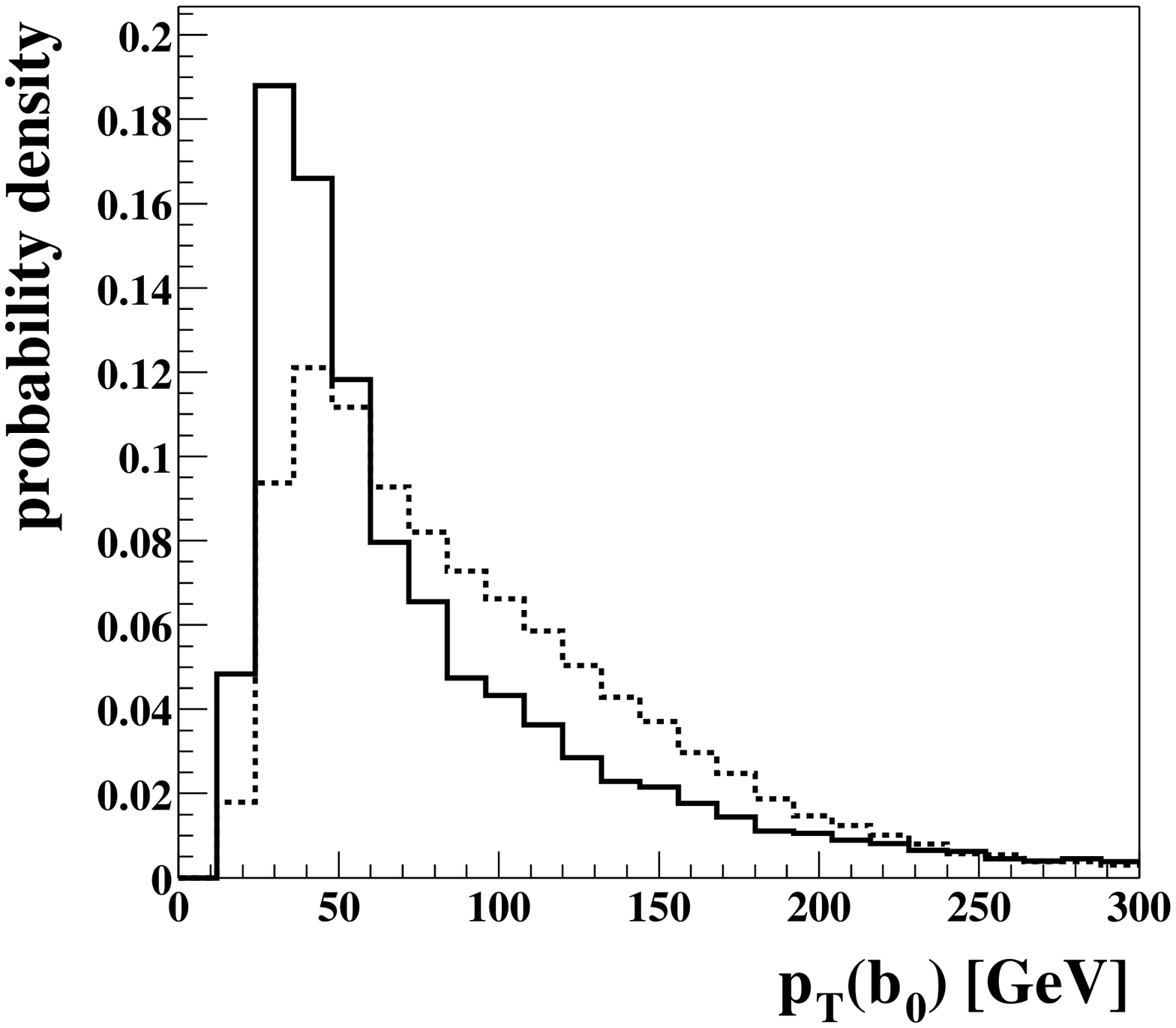,width=1.06\textwidth}
        \end{center}
      \end{minipage}
      \begin{minipage}[t]{.32\textwidth}
       \begin{center}
          \epsfig{file=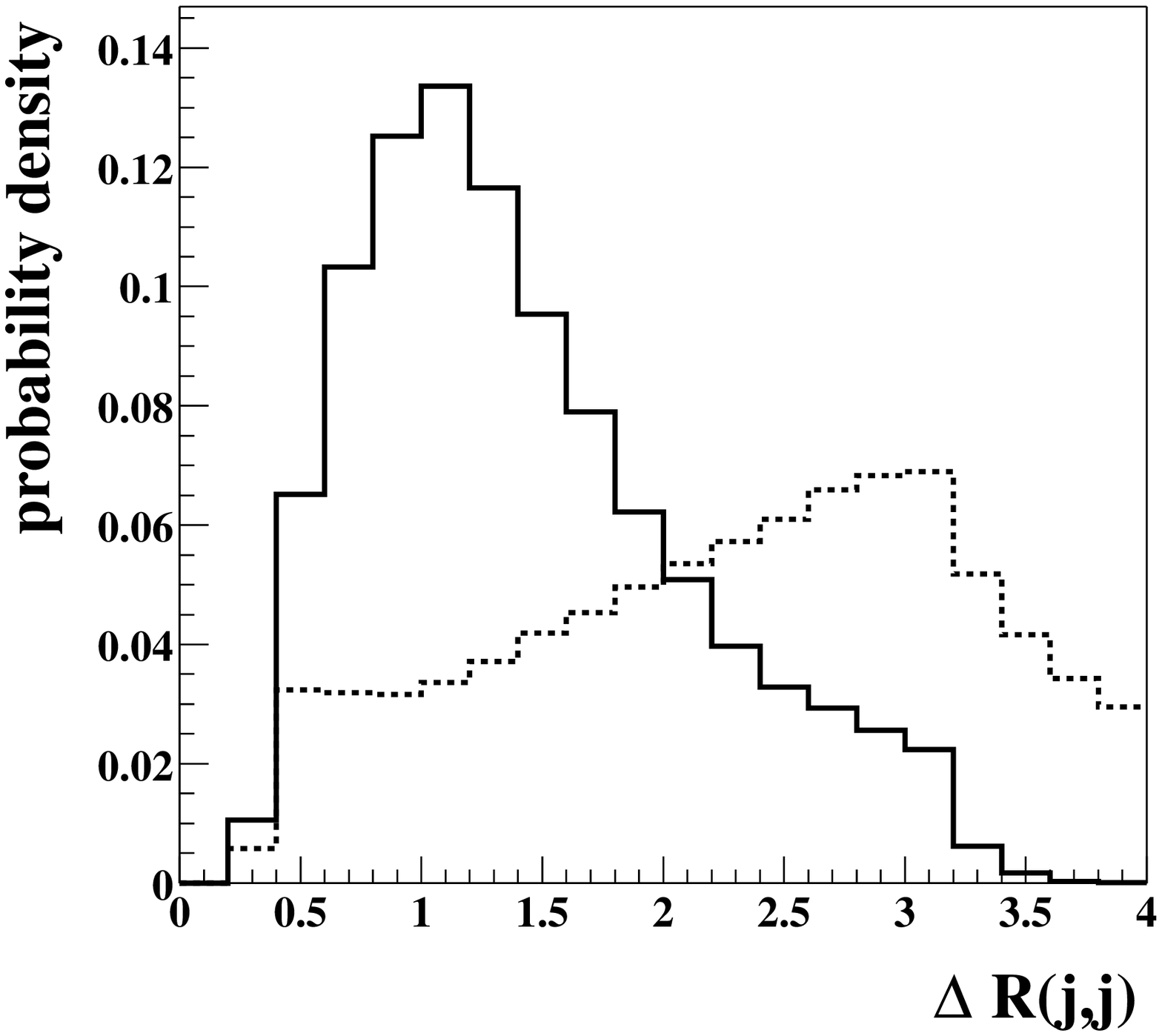,width=1.06\textwidth}
        \end{center}
      \end{minipage}
    \end{minipage}

    \begin{minipage}[t]{1\textwidth}
      \begin{minipage}[t]{.32\textwidth}
        \begin{center}
          \epsfig{file=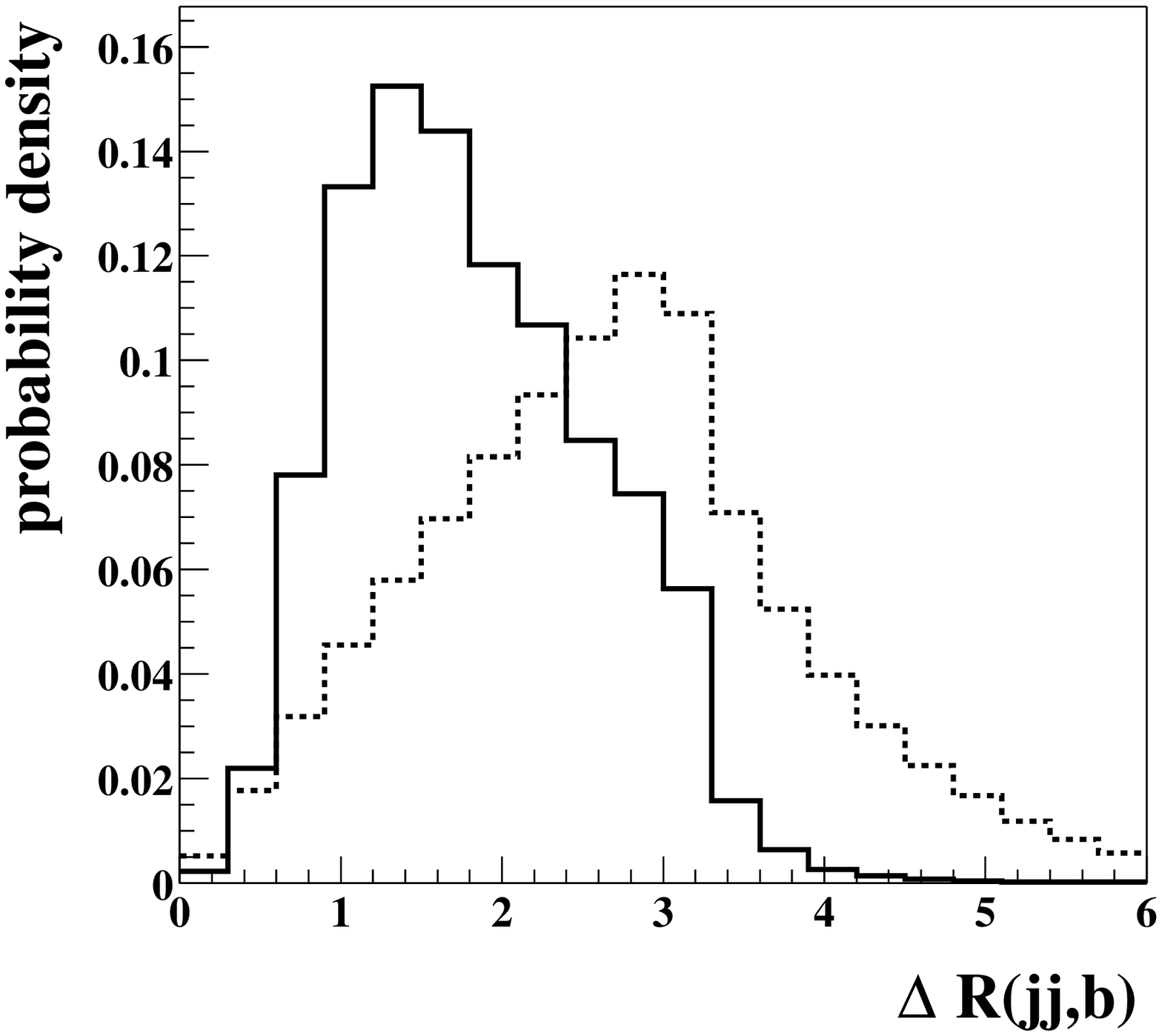,width=1.06\textwidth}
        \end{center}
      \end{minipage}
      \begin{minipage}[t]{.32\textwidth}
        \begin{center}
          \epsfig{file=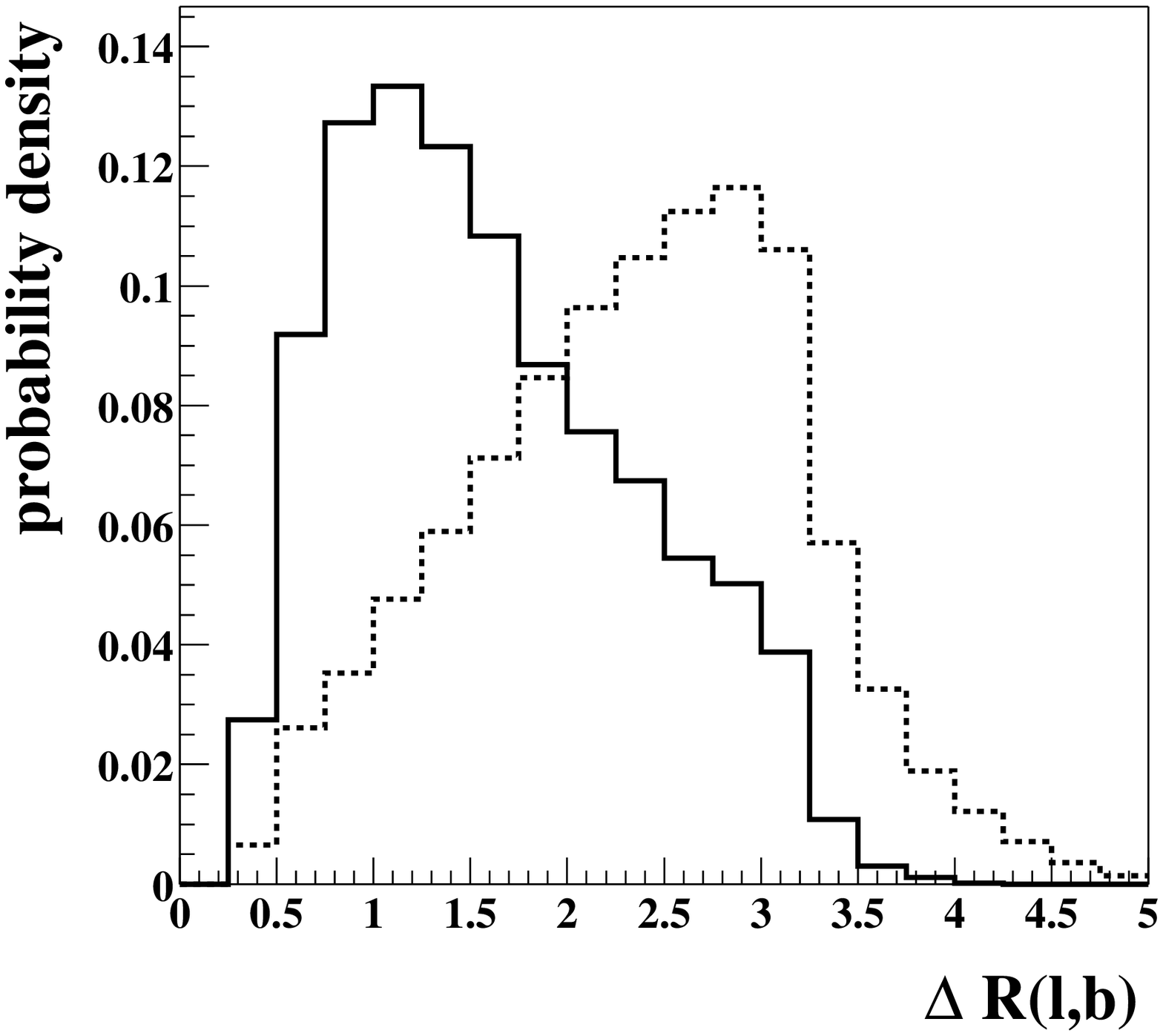,width=1.06\textwidth}
        \end{center}
      \end{minipage}
      \begin{minipage}[t]{.32\textwidth}
       \begin{center}
          \epsfig{file=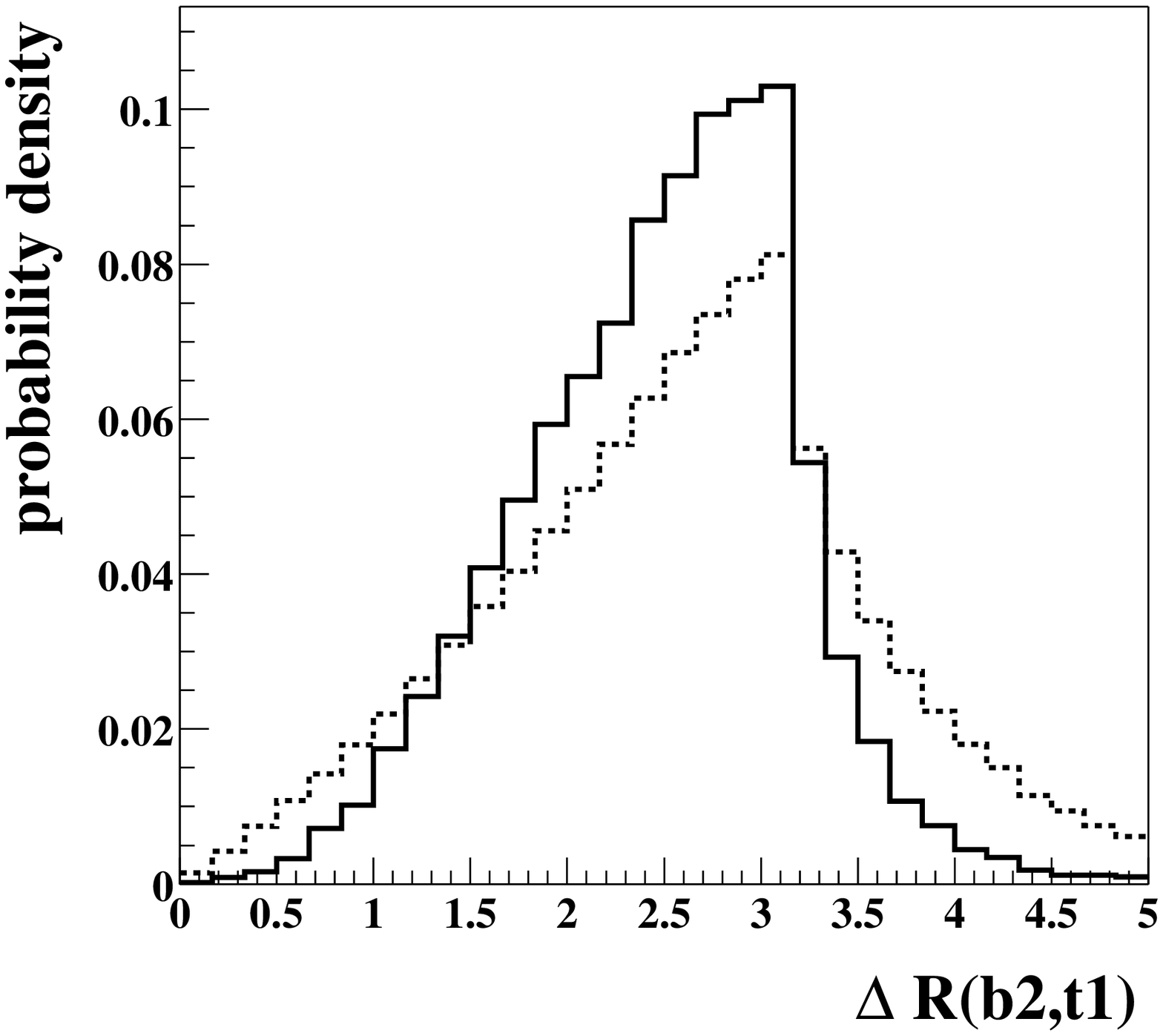,width=1.06\textwidth}
        \end{center}
      \end{minipage}
    \end{minipage}
\caption[]{\label{fig:combRefHist}\sl 
The probability density functions for the nine variables used in the
combinatorial likelihood for a charged Higgs mass of \mbox{$\mHpm=400\,\G$}. 
All distributions are normalised to unity including overflow bins.
Correct combinations are represented by a solid line, all wrong
combinations by a dashed line.
}  
\end{figure}

\begin{figure}[]
  \begin{center}
    \epsfig{file=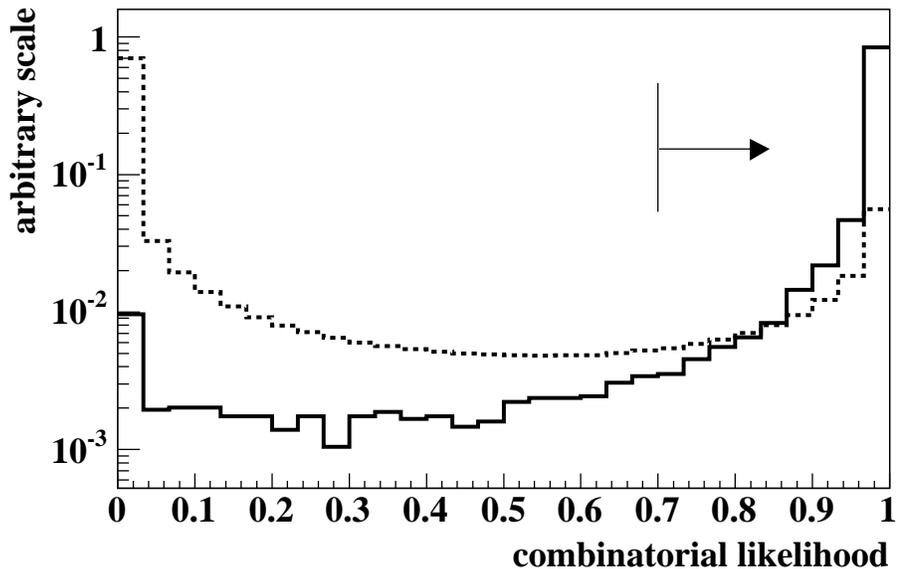,width=.8\textwidth}
  \end{center}
\caption[]{\label{fig:combRefLik}\sl 
The normalised combinatorial likelihood output distributions for the 
correct--combination class.
The correct (solid) and the wrong (dashed) combinations are
shown for a charged Higgs mass of \mbox{$\mHpm=400\,\G$}. 
Only combinations with a likelihood output larger than $0.7$ are accepted.
}  
\end{figure}

The combinatorial likelihood is based on the following $9$ variables, 
where the labelling of partons in the signal process is illustrated in 
Figure~\ref{fig:feynDiag}:
\begin{figure}[]
  \begin{center}
    \epsfig{file=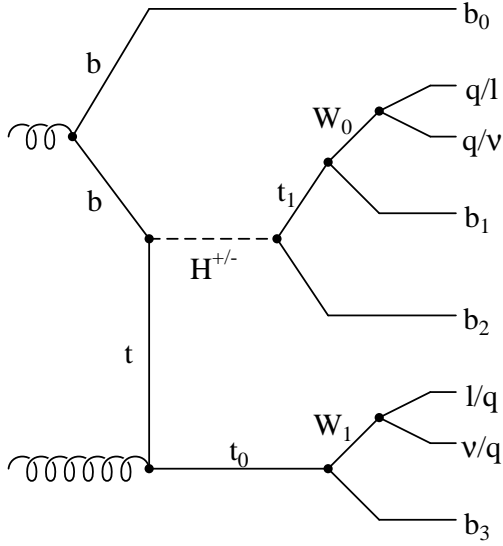,width=.4\textwidth}
  \end{center}
\caption[]{\label{fig:feynDiag}\sl 
One of the possible Feynman diagrams for the signal process 
\mbox{$gg\rightarrow tb\Hpm\rightarrow 
b\bar{b}b\bar{b}l\nu\bar{q}q'$},
illustrating the labelling of the partons adopted in this analysis.
}  
\end{figure}
\begin{enumerate}
\item\label{var:combRef1}
$m_{jj}$: the invariant mass of any two light jets. For the correct combination
this mass should be within the \Wpm\ mass peak around \mbox{$\mW=80.4\,\G$} 
whereas the distribution of
invariant masses of pairs of jets not originating from a \Wpm\ is
rather flat.
\item\label{var:combRef2}
$m_{jjb}$: the invariant mass of any two light jets and one of the
four $b$--jets. The correct combination should reproduce the top
mass. This variable aims at correctly reconstructing the hadronically
decaying top quark \thad.
\item\label{var:combRef3}
$m_{l{\nu}b}$: the invariant mass of the isolated lepton, one solution for the
neutrino reconstructed as described in \ref{sec:precuts} and one of
the $b$--jets.
This variable aims at reconstructing the leptonically decaying
top quark \tlep.
\item
$\pT(\btwo)$: the \pT\ of the $b$--jet assumed to originate 
from the charged Higgs decay.
\item
$\pT(\bzero)$: the \pT\ of the assumed companion $b$--jet 
produced in the \mbox{$gg\rightarrow tb\Hpm$} process.
\item
$\deltaR(j,j)$: \deltaR\ between any two light jets. 
Like $m_{jj}$
this variable helps to reconstruct \Whad.
\item
$\deltaR(jj,b)$: \deltaR\ between the sum of any 
two light jets and a $b$--jet. 
Like $m_{jjb}$ this variable is related to the reconstruction
of \thad.
\item
$\deltaR(l,b)$: \deltaR\ between the isolated lepton and a $b$--jet. 
Like $m_{l{\nu}b}$ this variable aims at reconstructing \tlep.
\item
$\deltaR(\btwo,\tone)$: \deltaR\ between the $b$--jet and the 
top quark candidate originating from the
charged Higgs decay. For the top quark candidate all possible modes of
reconstructing a top quark are considered. 
\end{enumerate}
The probability density functions for the nine variables used in the
combinatorial likelihood
are shown in Figure~\ref{fig:combRefHist} for a charged Higgs mass of
\mbox{$400\,\G$}.
Overflow bins are included in the normalisation of the histograms 
and used when calculating the combinatorial likelihood output.
The corresponding normalised likelihood distributions for the 
correct--combination class are shown in Figure~\ref{fig:combRefLik} for the
correct combination and all the	wrong ones.
As expected, the distribution corresponding to the correct 
combination peaks at $1$ whereas the distribution 
representing all the wrong combinations peaks at $0$. 
However, it is important to note the tail in the distribution
representing the wrong combinations up to
high likelihood values. The total number of combinations of
the reconstructed objects to reconstruct the event completely is given
by
\[ 
N=4! \times 
\left( \begin{array}{c} 
m\\
2  
\end{array} \right)
\times N_{\nu} \times 2, 
\]
where $m$ is the number of light jets in the event and 
$N_{\nu}$ is the number of solutions for the neutrino. 
The $4!$ represents the number of possibilities 
to order the four $b$--jets and the
factor of $2$ reflects the
possible associations of \Wlep\ and \Whad\ to the top quarks.
This number $N$ depends on the
number of light jets in the event and is generally quite large. Hence
the number of wrong combinations is large and the possibility of one
of those wrong combinations having a combinatorial likelihood
output higher than the correct combination is not negligible, thus 
reducing the probability of identifying the correct combination.

For each event the combination yielding the highest 
correct--combination likelihood is treated as the correct
combination. 
Nevertheless, if this selected combination yields a likelihood value below
$0.7$ the event is rejected. 
The efficiency of this cut varies between 
\mbox{$90\,\%$} for \mbox{$\mHpm=200\,\G$} and
\mbox{$95\,\%$} for \mbox{$\mHpm=800\,\G$} for the signal process and
is approximately \mbox{$85\,\%$} for the main 
\mbox{$gg\rightarrow t\bar{t}b\bar{b}$}
background.

\begin{figure}[]
  \begin{center}
    \epsfig{file=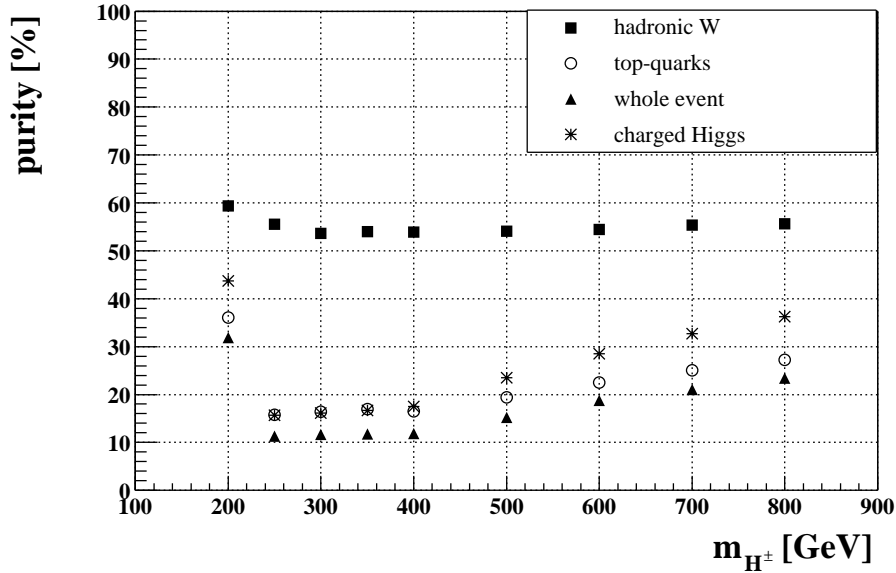,width=.8\textwidth}
  \end{center}
\caption[]{\label{fig:combRefPerf}\sl 
Purity of the combinatorial likelihood in reconstructing
the hadronically decaying \Wpm\ (squares), 
the two top quarks (circles), 
the charged Higgs boson (stars) 
and the whole event (triangles) over the
considered charged Higgs mass range.
}  
\end{figure}

The performance of the combinatorial likelihood is checked 
using the Monte Carlo truth information to associate the final state
partons with the reconstructed objects as described in section
\ref{sec:jetParton}. 
An event is classified as correctly reconstructed if the
four $b$--jets and the two light jets are correctly associated with their
corresponding final state partons and the correct lepton is
found to be isolated.
Some performance benchmarks of the combinatorial likelihood are shown 
in Figure~\ref{fig:combRefPerf} as a function of 
the charged Higgs boson mass.
The squares indicate the 
fraction of correctly reconstructed hadronically decaying \Wpm, 
referred to as purity in the following.
This purity does not depend strongly on the charged Higgs mass
and lies between \mbox{$53\,\%$} and \mbox{$60\,\%$}.
The purity of reconstructing the two top quarks 
is represented by the open circles.
Here the purity depends strongly on \mHpm\  
and rises to values above \mbox{$25\,\%$} only for charged Higgs masses
either very close to \mtop\ or for \mbox{$\mHpm\geq 700\,\G$}.
Similar behaviour can be seen for the purity of the 
charged Higgs boson (stars)
and the whole event reconstruction (triangles).
The only variables used in the combinatorial likelihood that
depend strongly on \mHpm\  are variables $4$ and $9$. 
The \pT\ of the $b$--jet from the charged Higgs decay
depends strongly on the mass of the charged Higgs boson and is well
separated from the average \pT\ of the other $b$--jets in the event only for
very small or quite large charged Higgs masses.
A similar statement applies to the distance \mbox{$\deltaR(\btwo,\tone)$}
between the top quark and the $b$--quark originating from the charged
Higgs decay. 
A light charged Higgs boson is produced with a sizable
boost and its decay products will have a small distance in 
$\deltaR$--space. 
On the other hand, a heavy charged Higgs boson is produced nearly at
rest and hence the distance between its decay products will be large.
The mass dependence of the performance of the combinatorial
likelihood is mainly determined by the mass dependence of 
variables $4$ and $9$.

It should be noted that in the mass region 
$\mHpm=250\,$--\mbox{$\,600\,\G$} 
the purity of the  
charged Higgs reconstruction 
does not exceed \mbox{$30\,\%$}.
As a consequence 
the reconstructed charged Higgs mass is substantially blurred
by the combinatorial background. 
Hence the detection of a clear mass peak in the reconstructed charged
Higgs mass distribution is difficult.
The correct reconstruction of the whole event is important for the
performance of the selection likelihood discussed below, 
since it relies on the correct association of reconstructed objects to 
the charged Higgs boson and the top quarks.
The ability to distinguish signal from background processes is 
already diminished by the imperfect performance of
the combinatorial likelihood.

\begin{figure}[]
    \begin{minipage}[t]{1\textwidth}
      \begin{minipage}[t]{.32\textwidth}
        \begin{center}
          \epsfig{file=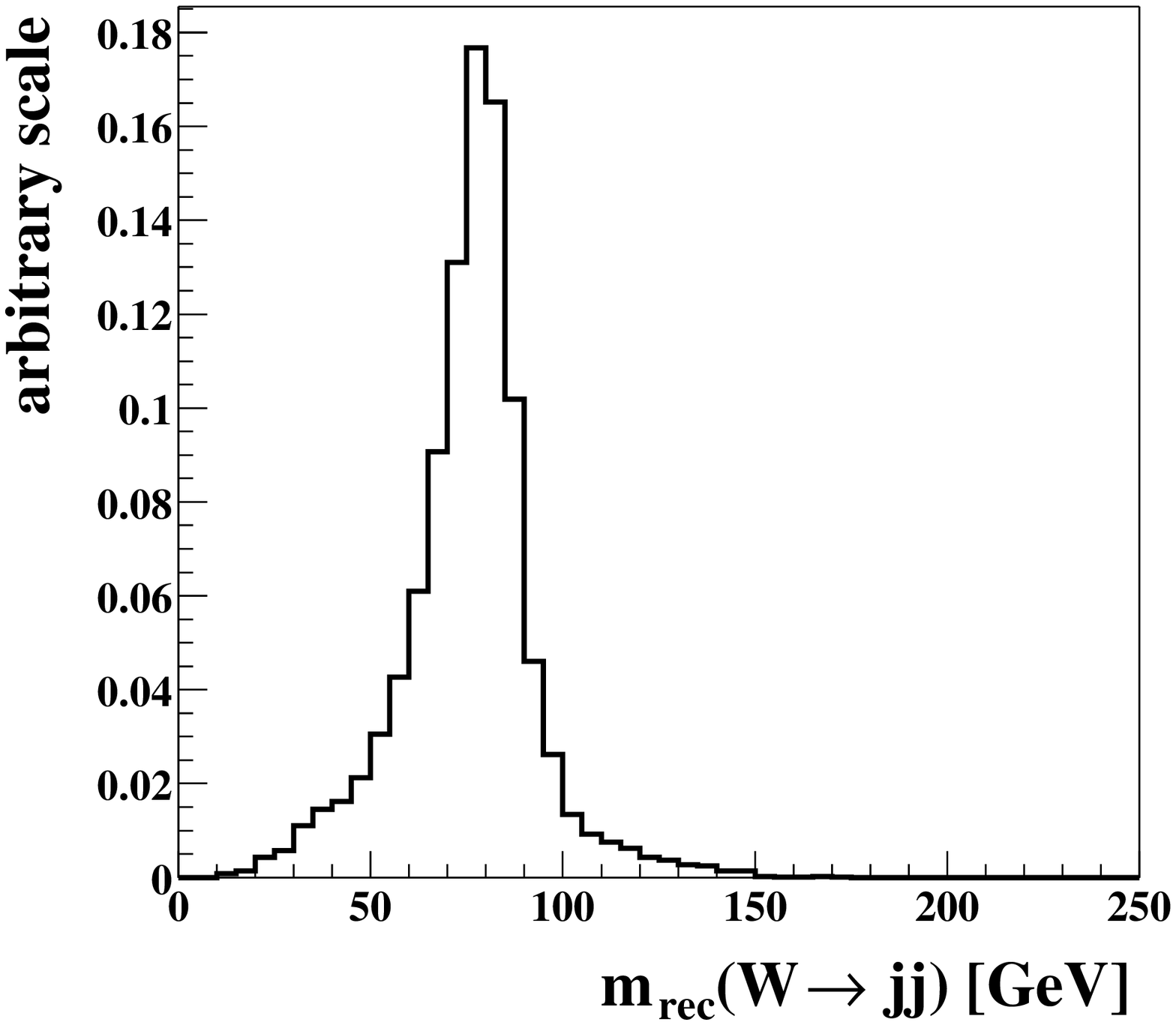,width=1.06\textwidth}
        \end{center}
      \end{minipage}
      \begin{minipage}[t]{.32\textwidth}
        \begin{center}
          \epsfig{file=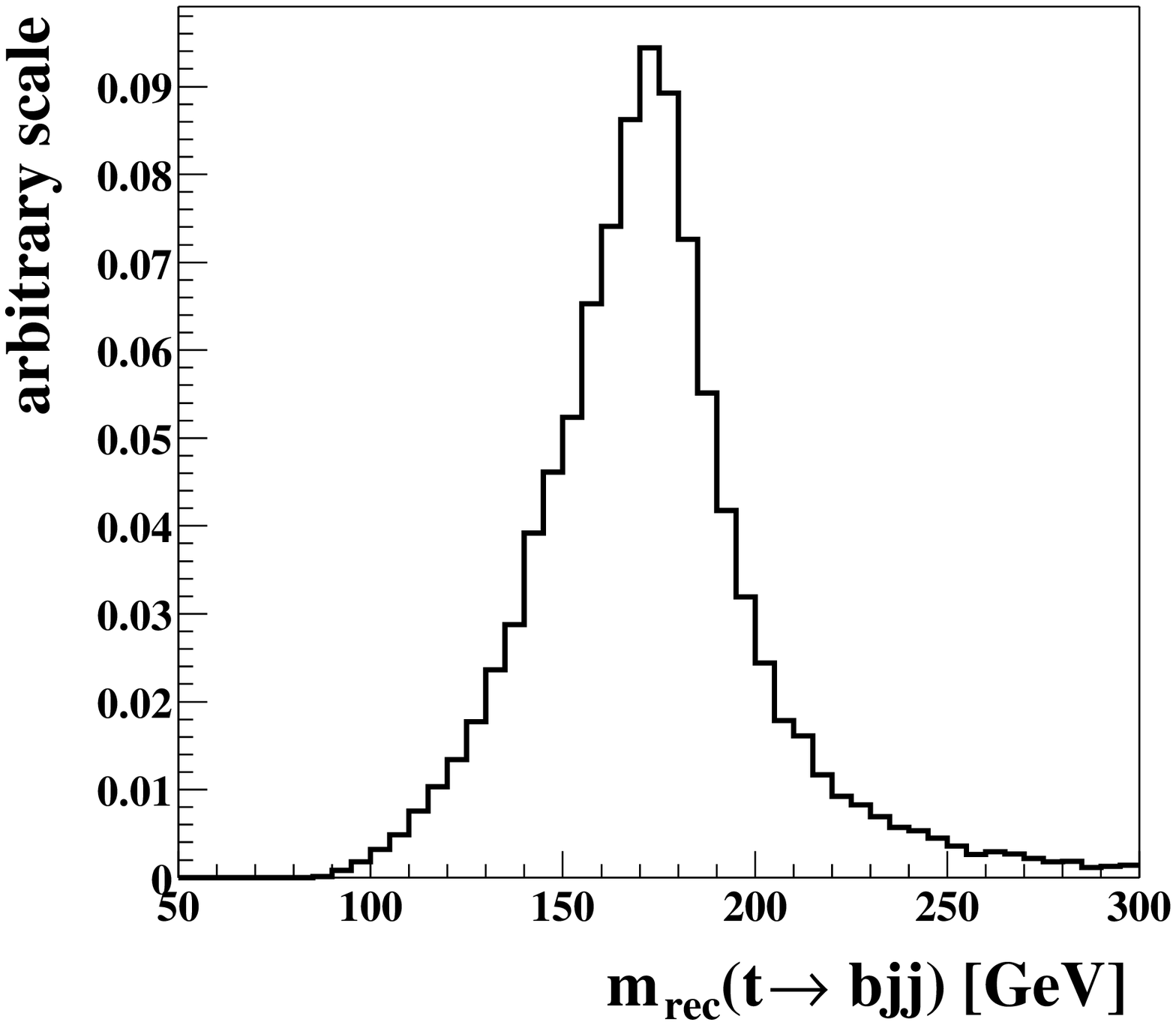,width=1.06\textwidth}
        \end{center}
      \end{minipage}
      \begin{minipage}[t]{.32\textwidth}
       \begin{center}
          \epsfig{file=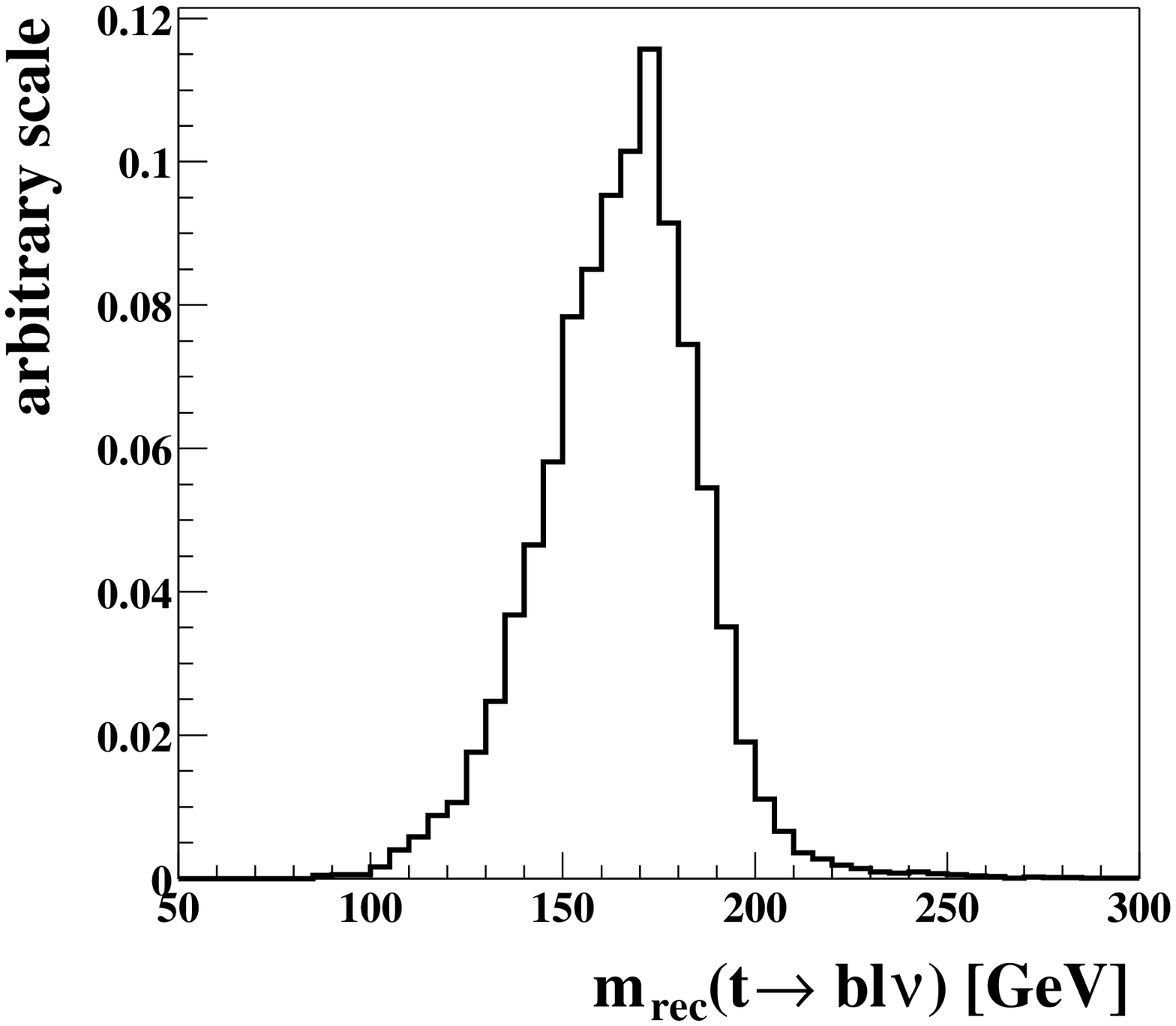,width=1.06\textwidth}
        \end{center}
      \end{minipage}
    \end{minipage}
\caption[]{\label{fig:recMasses}\sl 
Reconstructed masses for the hadronically decaying \Wpm\ and the hadronic
and leptonic top quark for events passing the cut on the
combinatorial likelihood. The signal is generated assuming
\mbox{$\mHpm=400\,\G$}.
}  
\end{figure}

\begin{figure}[]
    \begin{minipage}[t]{1\textwidth}
      \begin{minipage}[t]{.32\textwidth}
        \begin{center}
          \epsfig{file=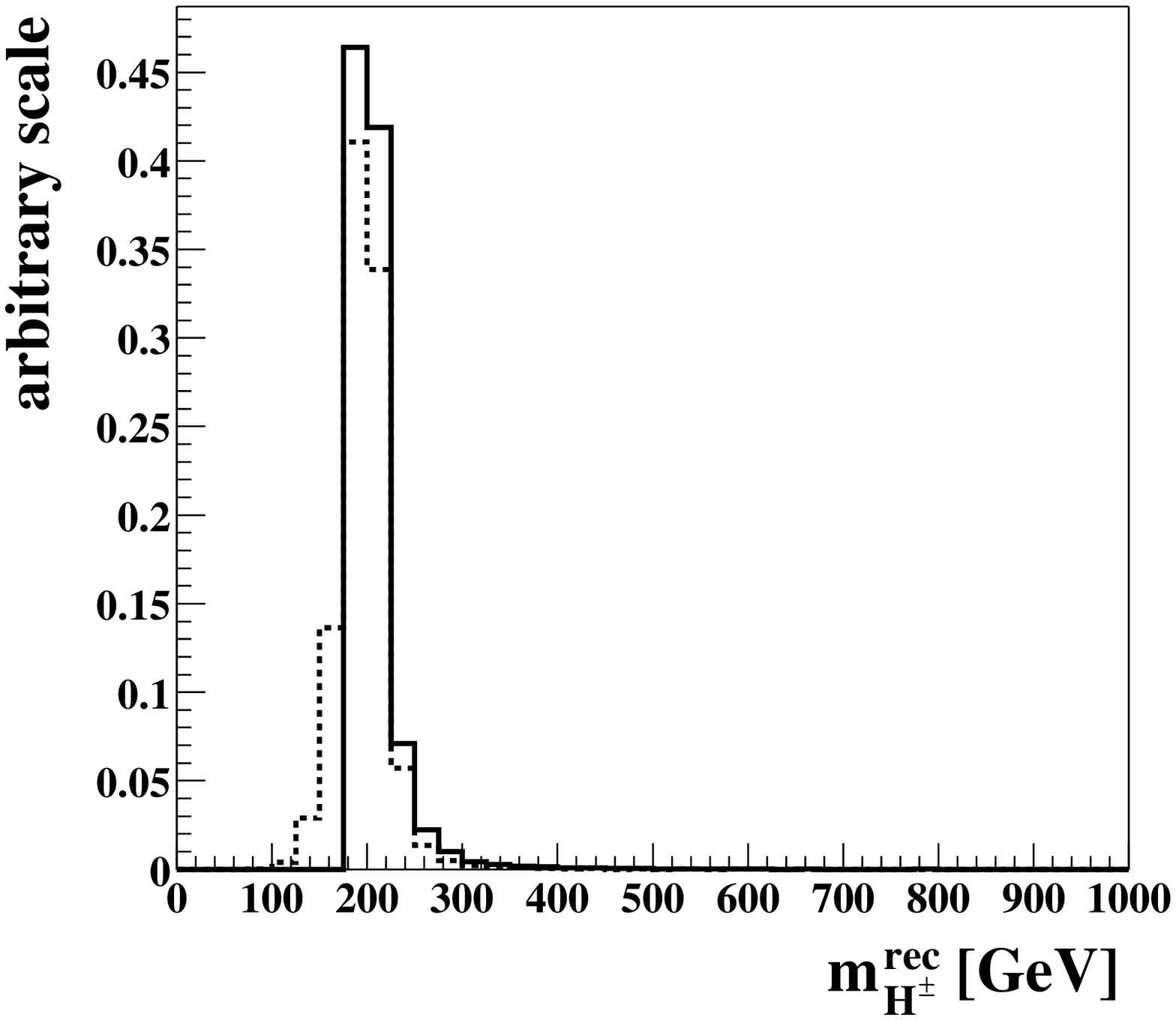,width=1.06\textwidth}
        \end{center}
      \end{minipage}
      \begin{minipage}[t]{.32\textwidth}
        \begin{center}
          \epsfig{file=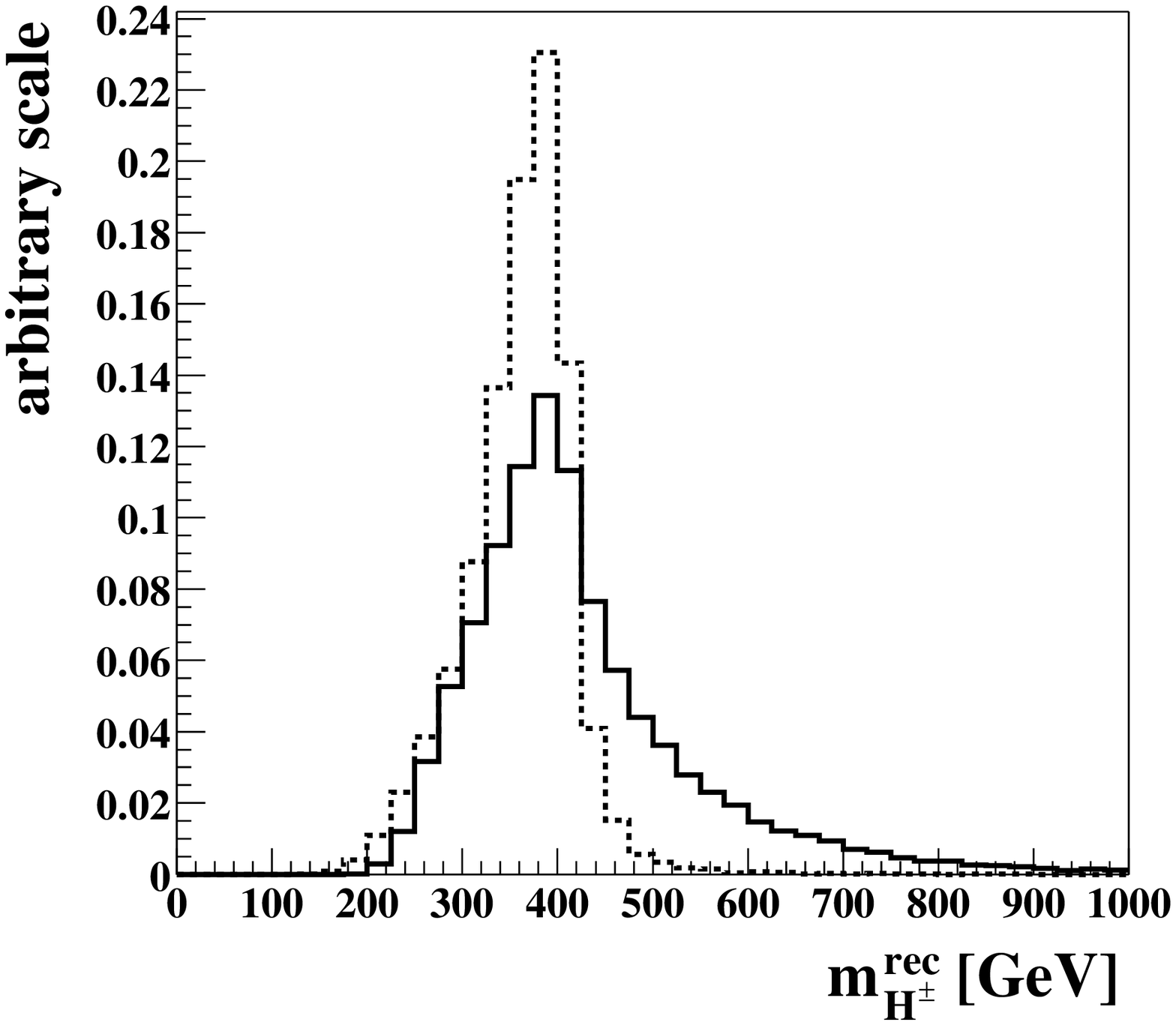,width=1.06\textwidth}
        \end{center}
      \end{minipage}
      \begin{minipage}[t]{.32\textwidth}
       \begin{center}
          \epsfig{file=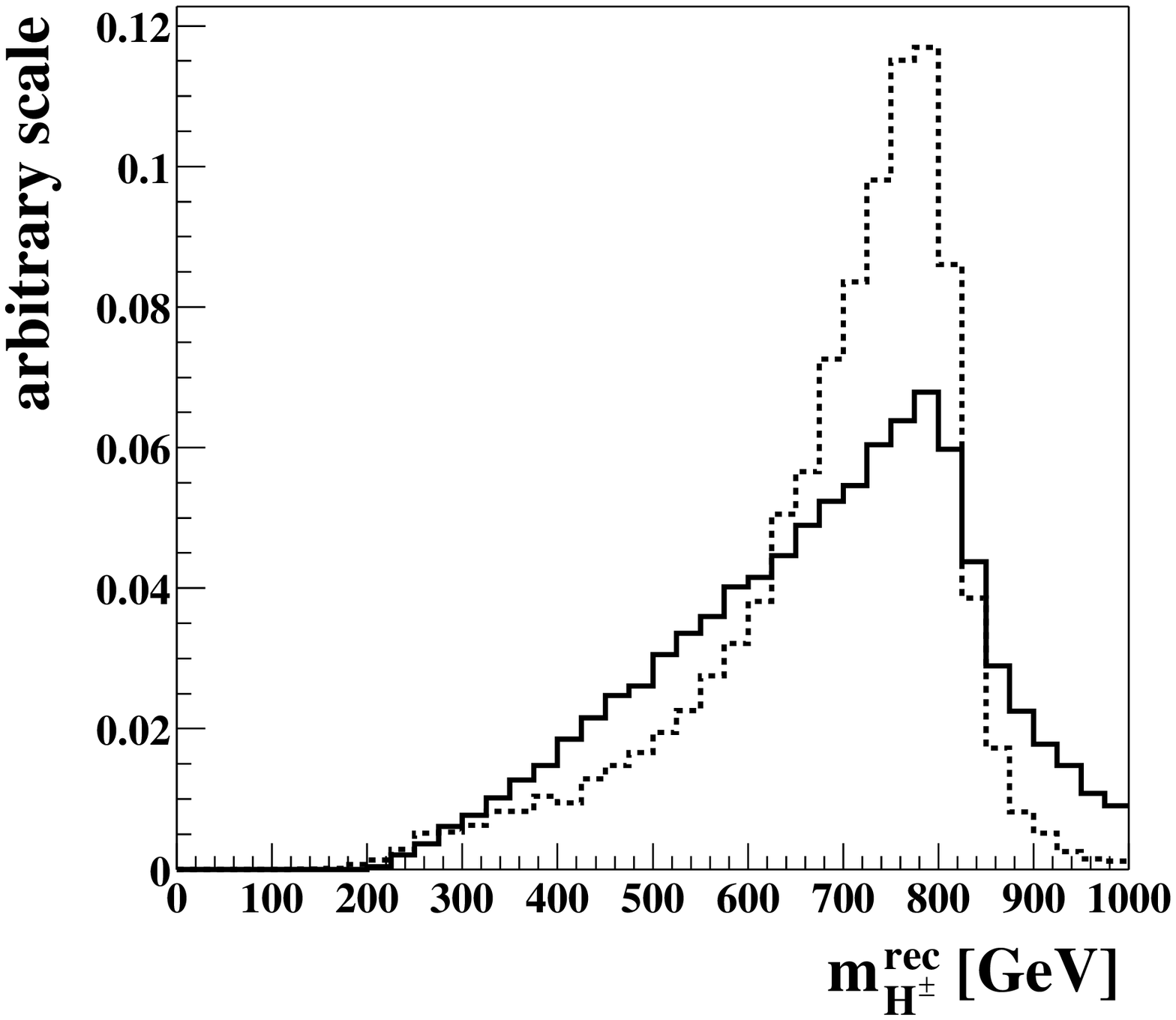,width=1.06\textwidth}
        \end{center}
      \end{minipage}
    \end{minipage}
\caption[]{\label{fig:recHmasses}\sl 
Reconstructed charged Higgs masses for $\mHpm=200$, $400$ and
$800\,\G$. The solid line shows the reconstructed mass as obtained
with the combinatorial likelihood. 
The dashed line illustrates the
charged Higgs mass resolution obtained when utilizing the
Monte Carlo truth information to match the tree--level partons to the 
reconstructed objects.
}  
\end{figure}

Before combining a \Wpm\ and a $b$--quark to form a reconstructed
top quark, the \Wpm\ 4--momentum is scaled to reproduce
\mbox{$m_{\Wpm}=80.4\,\G$}. 
The reconstructed hadronic \Wpm\ and the two reconstructed top masses for 
events passing the cut of $0.7$ on the combinatorial likelihood output 
are shown in Figure~\ref{fig:recMasses}. 
Reconstructed charged Higgs masses are shown for \mbox{$\mHpm=200$},
$400$ and \mbox{$800\,\G$} in Figure~\ref{fig:recHmasses}. The solid
line represents the reconstructed mass obtained with the combinatorial
likelihood. To predict the detector performance and to illustrate the
effect of the combinatorial background the same distributions are also
shown using the Monte Carlo truth information to select the correct
combination of reconstructed objects as the dashed lines. The effect
of the combinatorial background is clearly visible as a tail in the
reconstructed charged Higgs mass distribution, especially toward
higher reconstructed masses. 
For higher charged Higgs masses detector effects become 
more prominent.  
Even when the MC truth information is included 
a large tail toward lower reconstructed masses develops.

\subsection{The Selection Likelihood}
\label{sec:selLik}
\begin{figure}[]
    \begin{minipage}[t]{1\textwidth}
      \begin{minipage}[t]{.45\textwidth}
        \begin{center}
          \epsfig{file=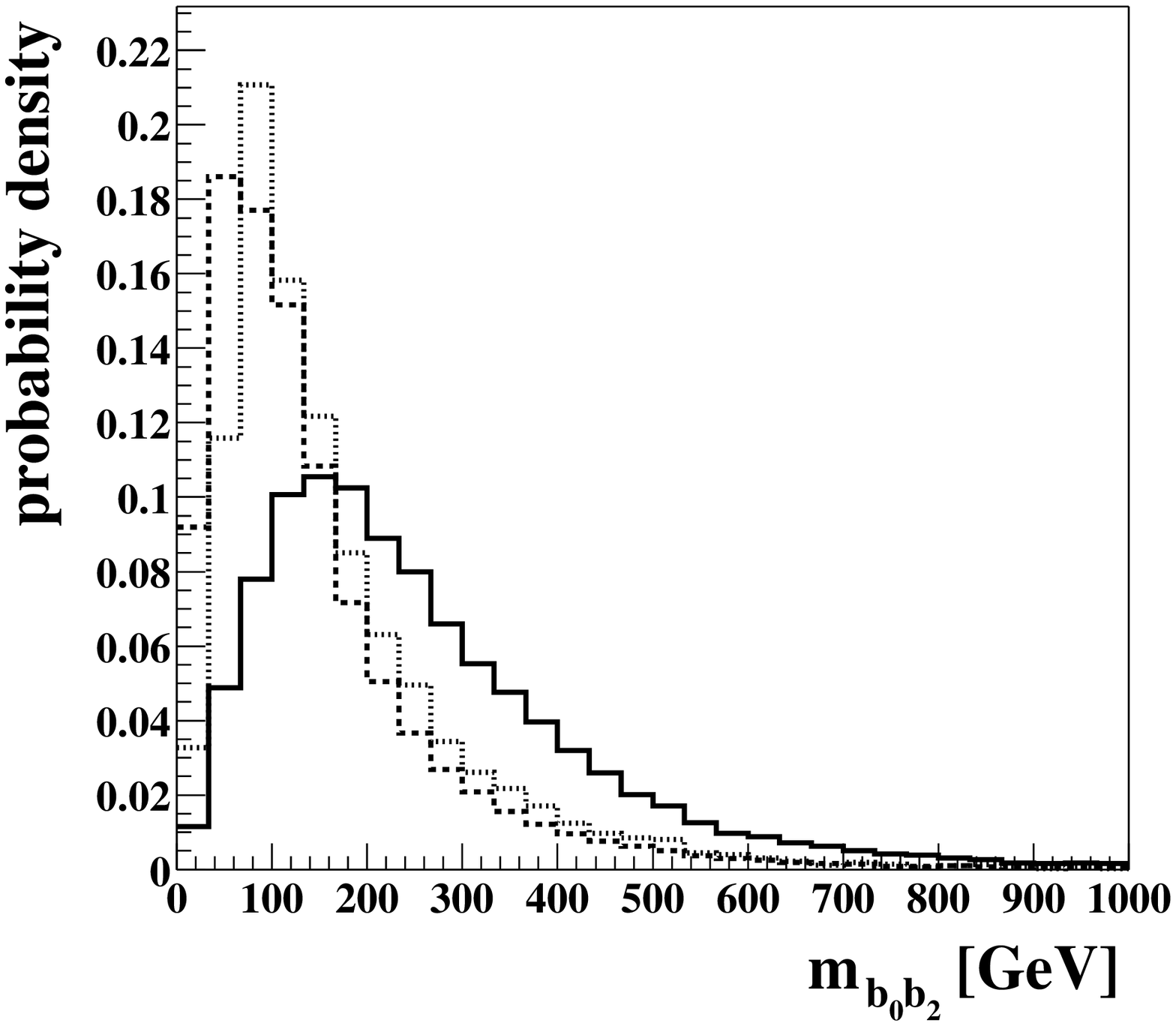,width=1.06\textwidth}
        \end{center}
      \end{minipage}
      \begin{minipage}[t]{.45\textwidth}
        \begin{center}
          \epsfig{file=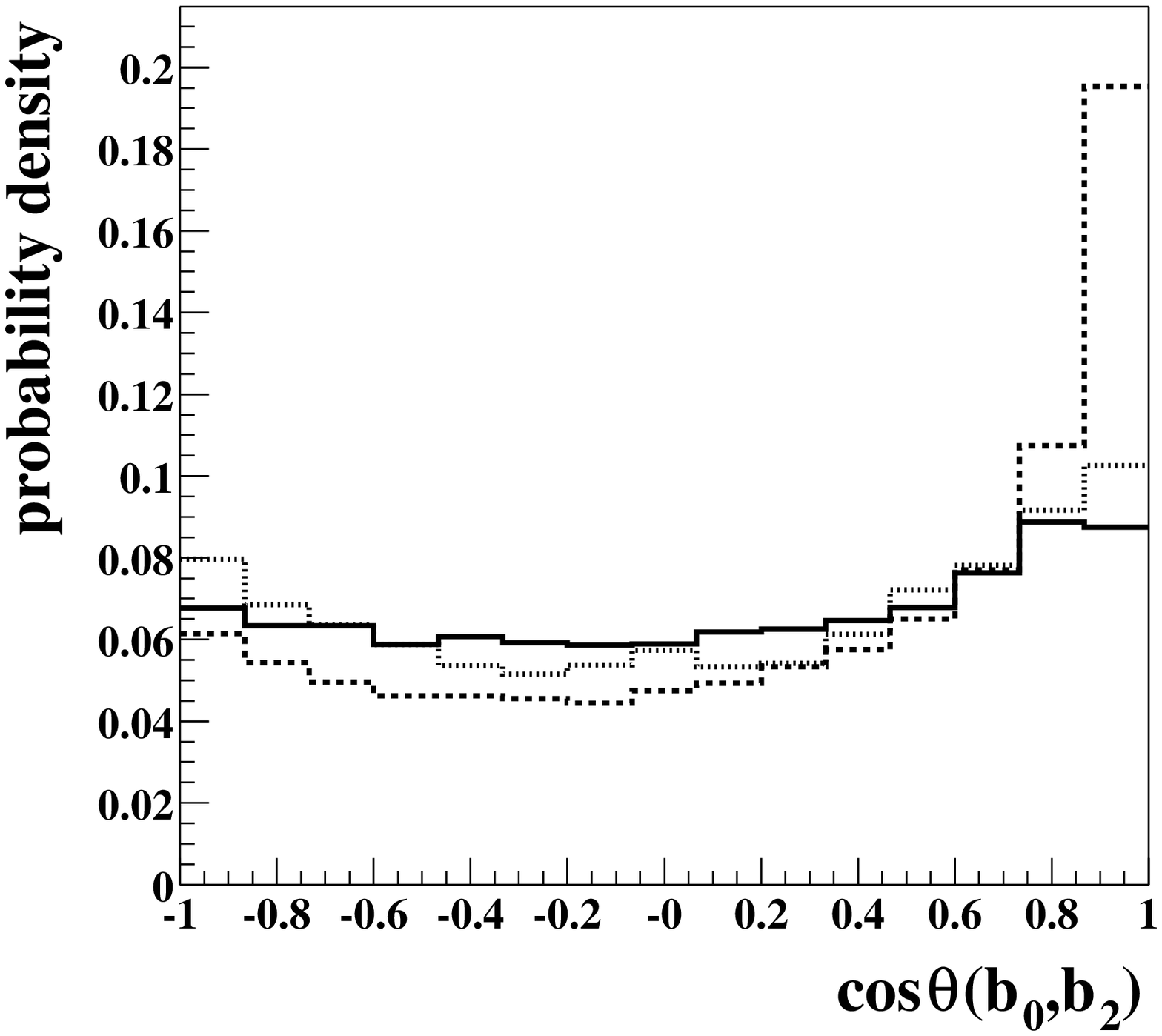,width=1.06\textwidth}
        \end{center}
      \end{minipage}
    \end{minipage}

    \begin{minipage}[t]{1\textwidth}
      \begin{minipage}[t]{.45\textwidth}
        \begin{center}
          \epsfig{file=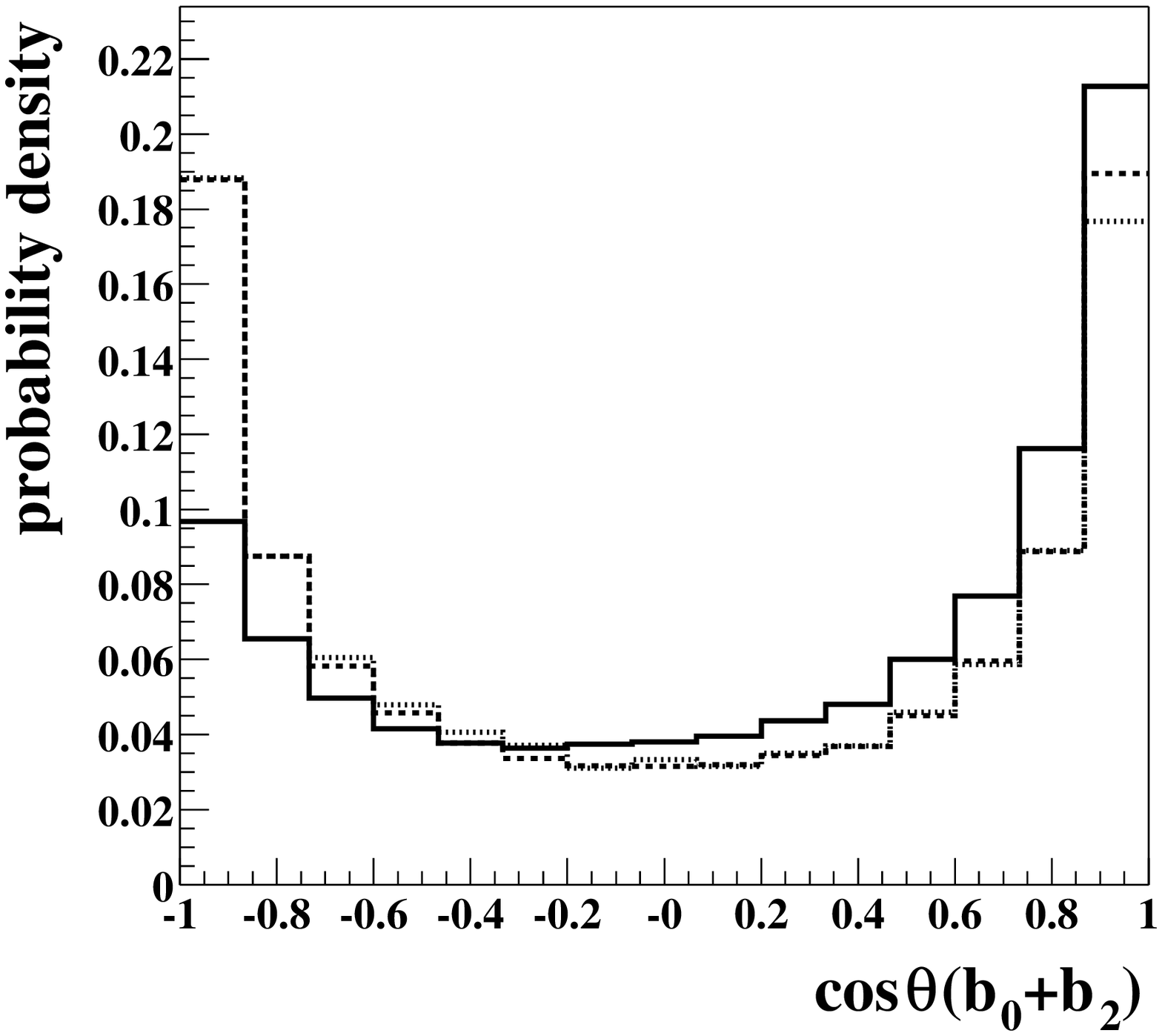,width=1.06\textwidth}
        \end{center}
      \end{minipage}
      \begin{minipage}[t]{.45\textwidth}
        \begin{center}
          \epsfig{file=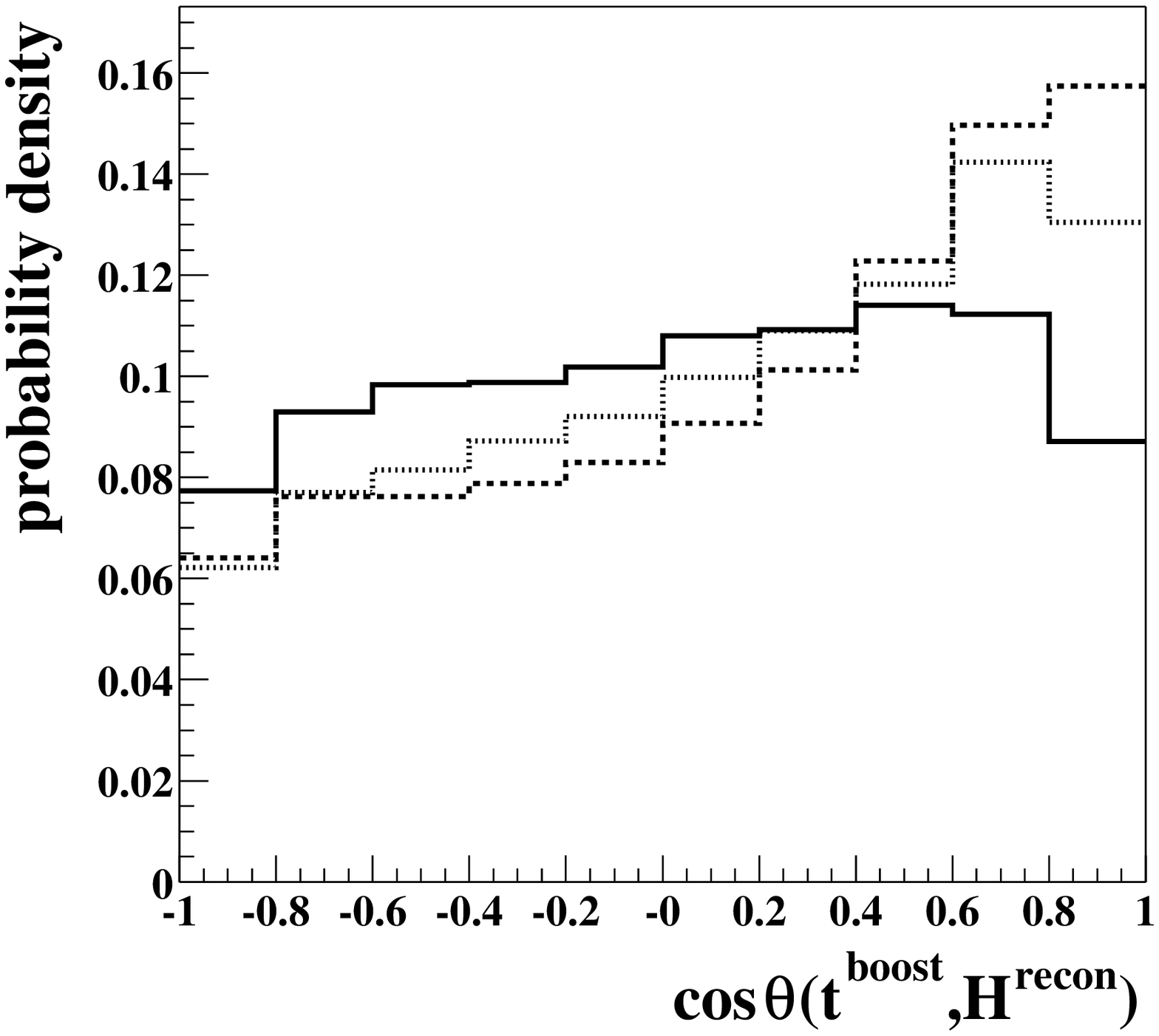,width=1.06\textwidth}
        \end{center}
      \end{minipage}
    \end{minipage}
\caption[]{\label{fig:selRefHist}\sl 
The probability density functions for the four variables used in the
selection likelihood for a charged Higgs mass of \mbox{$\mHpm=600\,\G$}. 
All distributions are normalised to unity.
For each variable the distributions corresponding to 
the \mbox{$gg\rightarrow tb\Hpm$} signal process (solid),
the \mbox{$gg / qq\rightarrow t\bar{t}b\bar{b}$} background process (dashed) and 
the electroweak \mbox{$gg\rightarrow Z/\gamma/W\rightarrow t\bar{t}b\bar{b}$}
process (dotted)
are shown.
}  
\end{figure}

To enhance the signal and suppress the \SM\ background 
a second likelihood selection is implemented.  
The selection likelihood distinguishes three classes of events:
1)~the \mbox{$gg\rightarrow tb\Hpm$} signal process,
2)~the \mbox{$gg / qq\rightarrow t\bar{t}b\bar{b}$} background, and 
3)~the \mbox{$gg\rightarrow Z/\gamma/W\rightarrow t\bar{t}b\bar{b}$}
background process and is implemented using the same formalism as 
described in section~\ref{sec:combLik}.
It exploits differences between the distributions of the signal and the
\SM\ backgrounds in the following four variables:
\begin{enumerate}
\item
$m_{b_{0}b_{2}}$: the invariant mass of the two $b$--jets not
originating from a top quark decay. In the signal events one of the
jets originates from a heavy charged Higgs boson whereas in the
background processes both $b$--jets originate from gluon
splitting. Hence the invariant mass of the two jets is expected to be
lower in background than in signal events.
\item
$\cos\theta(b_0,b_2)$: the cosine of the angle between the two $b$--jets
not originating from a top quark decay. Since the two $b$--jets
originate from gluon splitting for the background processes they
are expected to be collinear whereas the distribution should be flat
for the signal process.
\item
$\cos\theta(b_{0}+b_{2})$: the cosine of the azimuthal angle of the
$b_{0}+b_{2}$ jet system.
\item
$\cos\theta(t_{\mathrm{boost}},\Hpm_{\mathrm{recon}})$: the cosine of
the angle between the reconstructed charged Higgs boson momentum and
the reconstructed top quark associated with its decay, where the
reconstructed top quark 4--momentum is boosted into the charged Higgs
boson rest frame.
\end{enumerate}
The corresponding probability density distributions are shown in
Figure~\ref{fig:selRefHist} for \mbox{\mHpm=600\,\G}.

Variables related to transverse momenta or invariant
masses of jet systems tend to shift the background peak in the reconstructed 
charged Higgs mass distribution towards the nominal charged Higgs
mass. 
Therefore only variables involving angular correlations are used in the
selection likelihood. The only exception is $m_{b_{0}b_{2}}$ 
for which it has been demonstrated
that no such shift occurs for 
cuts of less than~$0.4$ on the resulting selection likelihood.

\begin{figure}[]
  \begin{center}
    \epsfig{file=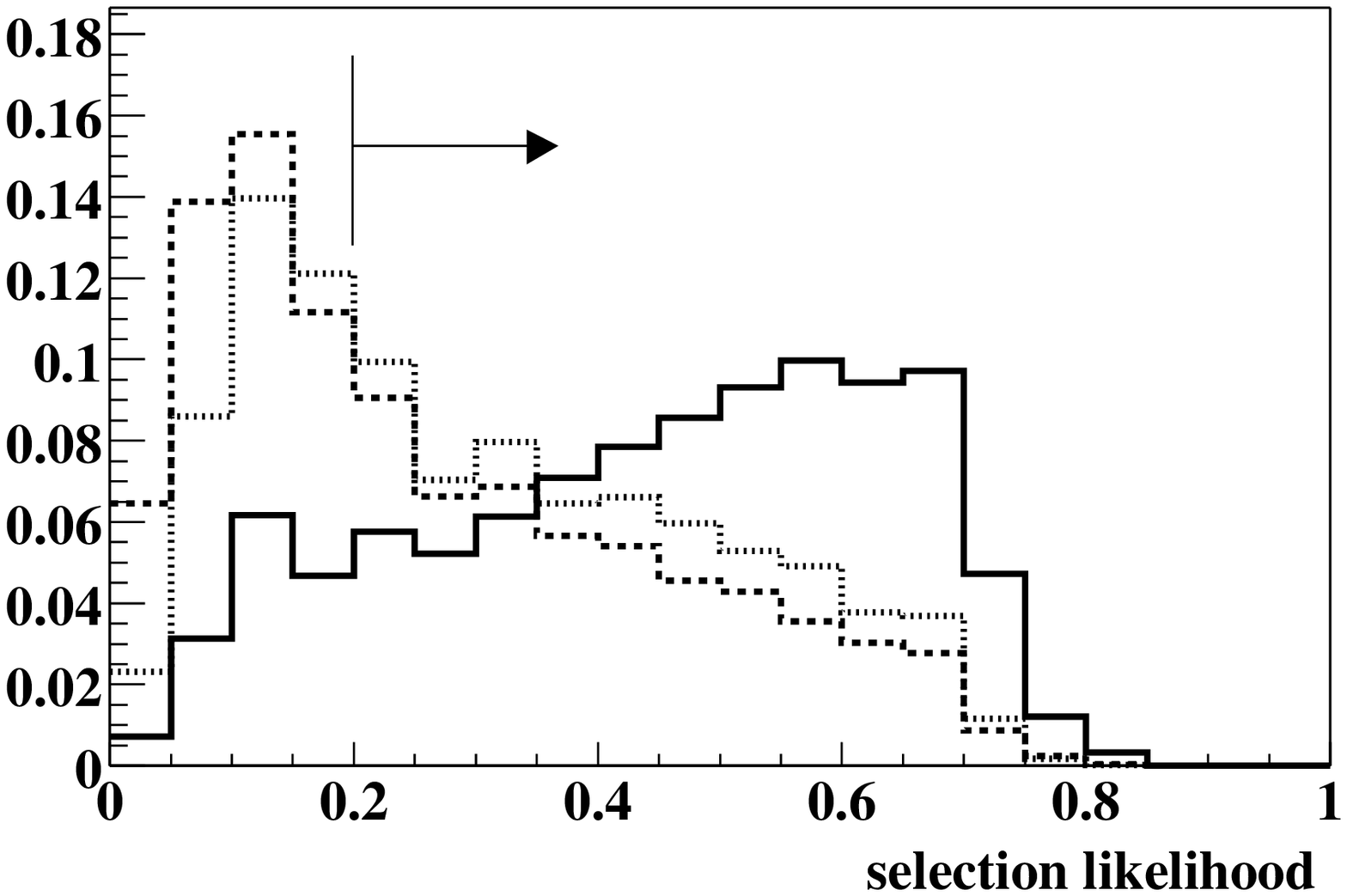,width=.8\textwidth}
  \end{center}
\caption[]{\label{fig:selLikDist}\sl 
The normalised selection likelihood output distributions for the 
signal class.
Distributions for 
the \mbox{$gg\rightarrow tb\Hpm$} signal process (solid),
the \mbox{$gg / qq\rightarrow t\bar{t}b\bar{b}$} background process (dashed) and 
the electroweak \mbox{$gg\rightarrow Z/\gamma/W\rightarrow t\bar{t}b\bar{b}$}
process (dotted)
are shown for a charged Higgs mass of \mbox{$\mHpm=600\,\G$}. 
Only events yielding a selection likelihood output larger than
$0.2$ are selected.
}  
\end{figure}

\begin{figure}[]
    \begin{minipage}[t]{1\textwidth}
      \begin{minipage}[t]{.45\textwidth}
        \begin{center}
          \epsfig{file=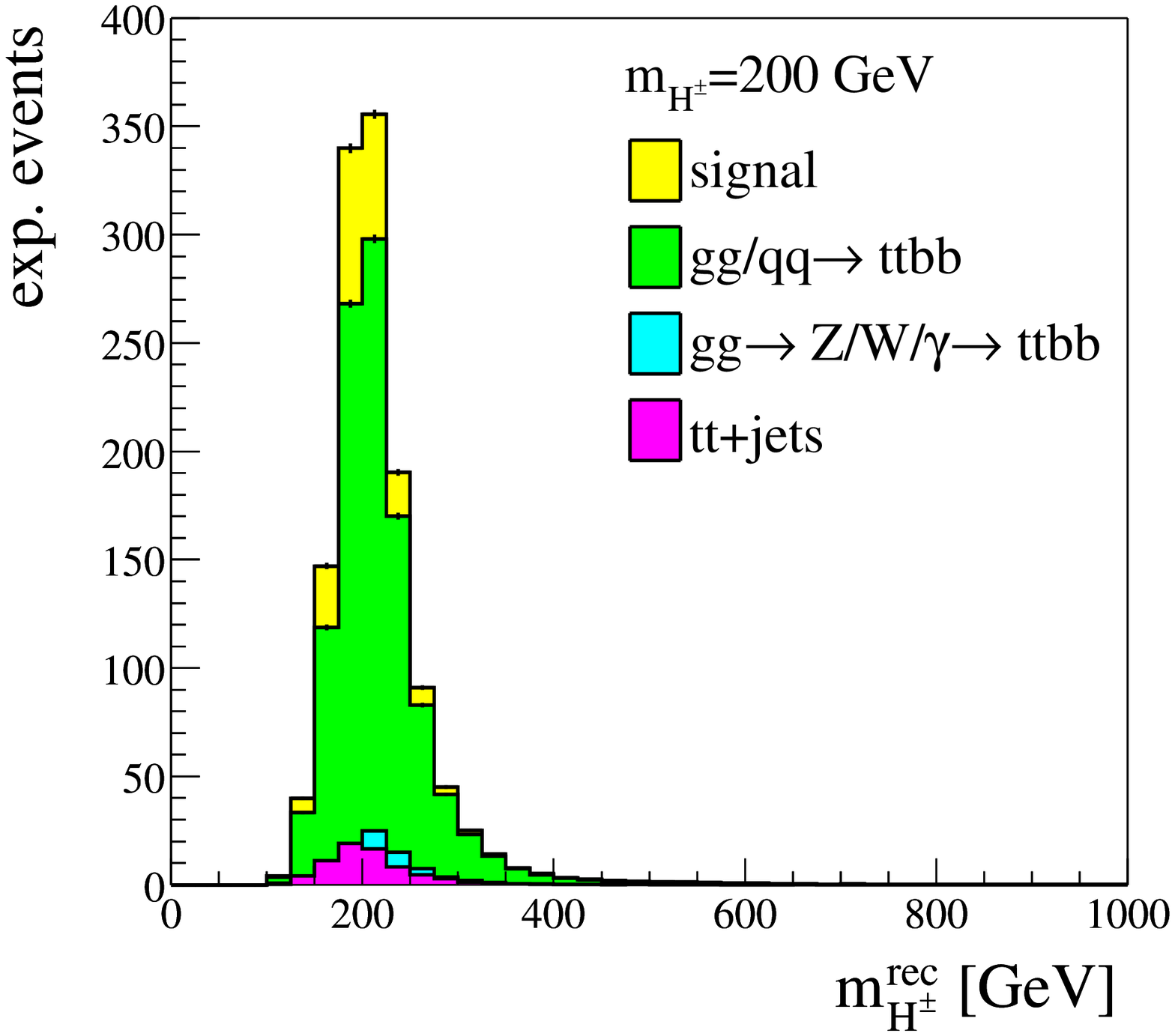,width=1.06\textwidth}
        \end{center}
      \end{minipage}
      \begin{minipage}[t]{.45\textwidth}
        \begin{center}
          \epsfig{file=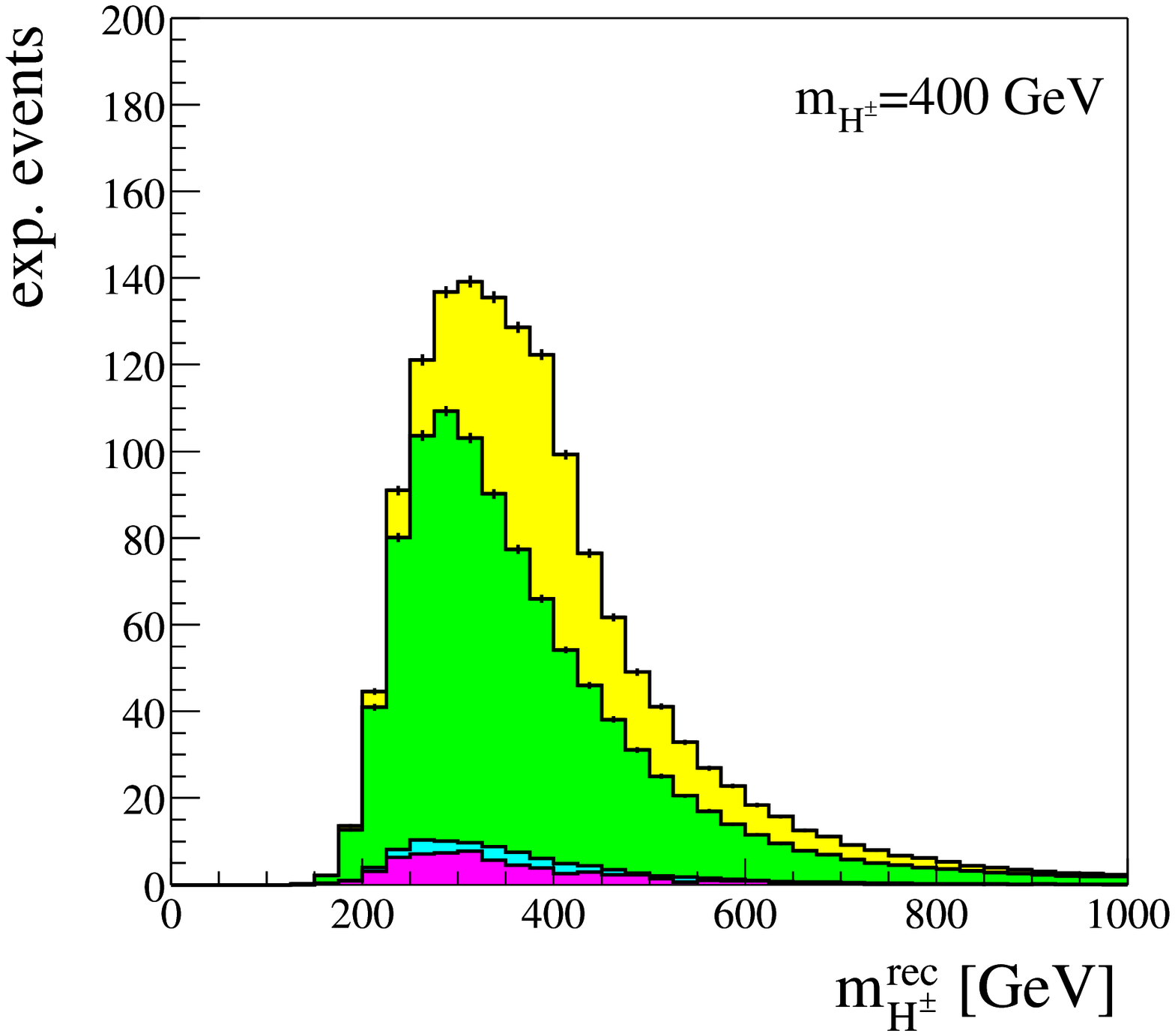,width=1.06\textwidth}
        \end{center}
      \end{minipage}
    \end{minipage}

    \begin{minipage}[t]{1\textwidth}
      \begin{minipage}[t]{.45\textwidth}
        \begin{center}
          \epsfig{file=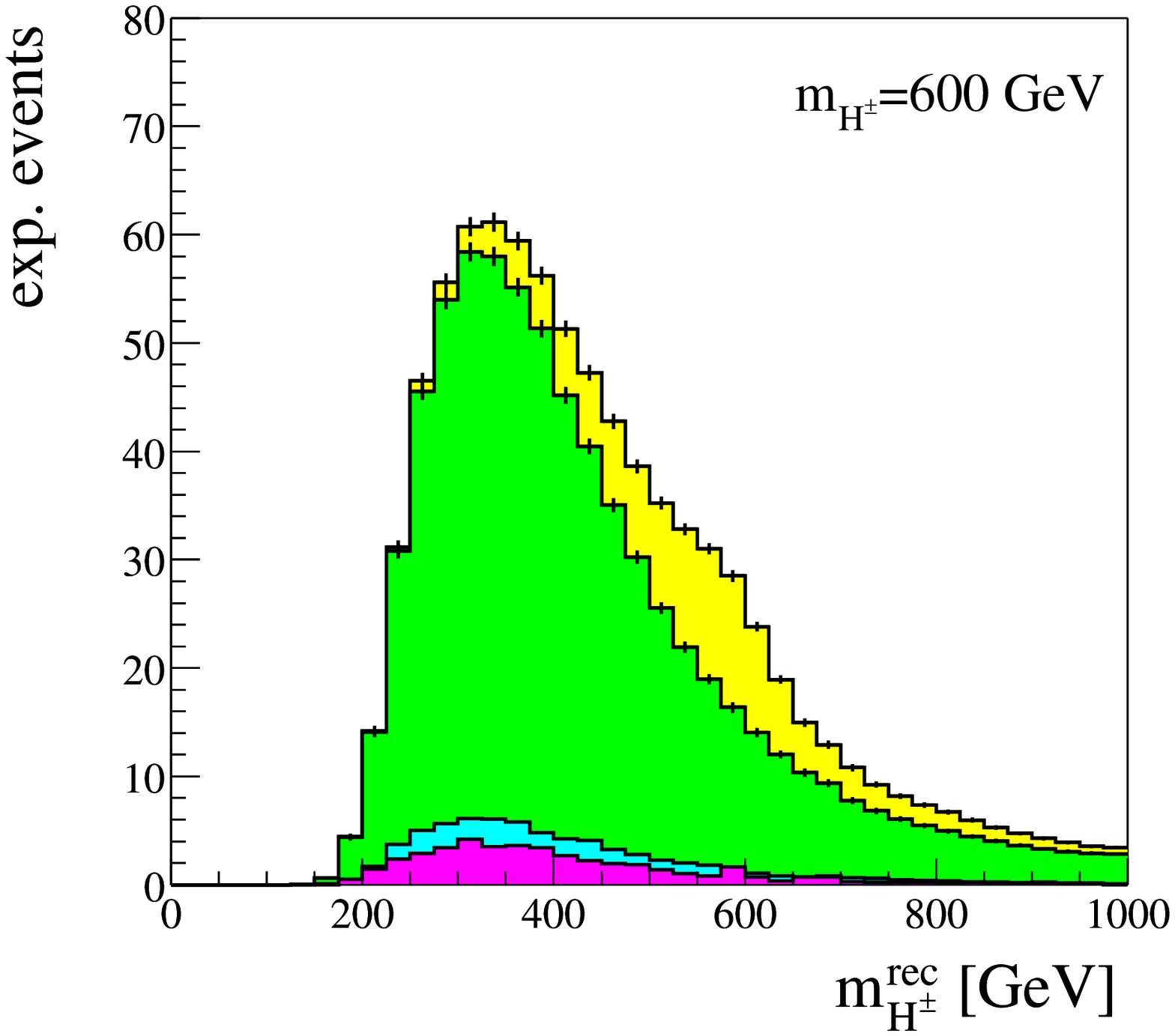,width=1.06\textwidth}
        \end{center}
      \end{minipage}
      \begin{minipage}[t]{.45\textwidth}
        \begin{center}
          \epsfig{file=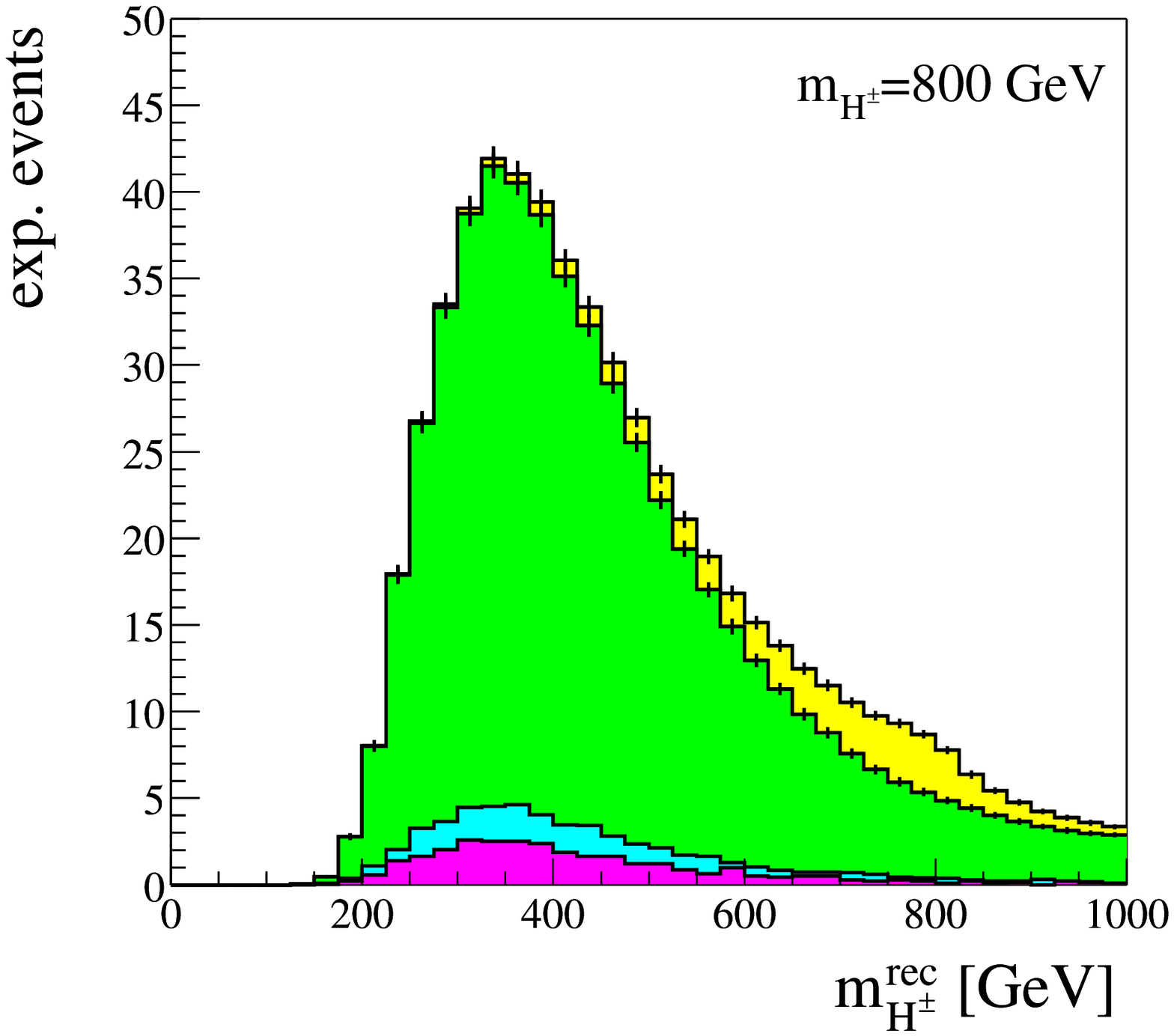,width=1.06\textwidth}
        \end{center}
      \end{minipage}
    \end{minipage}
\caption[]{\label{fig:reconMasses}\sl 
Reconstructed charged Higgs masses for $\mHpm = 200$, $400$, $600$ 
and $800\,\G$ and $\tanb=80$. The histograms are normalised to the
expected event rate for \mbox{${\cal L}=30\,\ifb$}.
The \mbox{$gg\rightarrow tb\Hpm$} signal process (yellow) is shown on top of
the \SM\ backgrounds
\mbox{$gg / qq\rightarrow t\bar{t}b\bar{b}$} (green),
\mbox{$gg\rightarrow Z/\gamma/W\rightarrow t\bar{t}b\bar{b}$} (light blue)
and \mbox{$t\bar{t}+\mathrm{jets}$} (pink).
}  
\end{figure}

The normalised selection likelihood output distributions for the
signal class and assuming \mbox{$\mHpm=600\,\G$} are shown in
Figure~\ref{fig:selLikDist}.
These distributions are calculated for each charged Higgs mass 
under consideration.
Signal events are separated from the \SM\ background by 
selecting only those events yielding a selection likelihood 
output larger than $0.2$. 
This optimal cut value is found by varying the cut on the likelihood output
in steps of $0.05$ and requiring the method to yield the 
highest discovery potential over the
selected range of charged Higgs masses.
Figure~\ref{fig:reconMasses} shows the resulting reconstructed charged
Higgs mass distribution for a choice of charged Higgs masses. 
Here an integrated luminosity of \mbox{${\cal L}= 30\,\ifb$} and 
\mbox{$\tanb=80$} is assumed.
Only a slight shift of the peak in the reconstructed 
charged Higgs boson mass for the
backgrounds is observed for growing $\mHpm$. 
For \mbox{$\mHpm\gtrsim 400\,\G$} the peaks in the reconstructed
charged Higgs mass for the signal and the background processes 
can be separated.

All events within a mass window of \mbox{$\pm 100\,\G$} around the
nominal charged Higgs mass are selected. 
The width of the mass window has little
influence on the discovery potential and hence 
is not optimised.
The number of selected signal and background events is then treated
like in a simple counting experiment and the Poisson significance is
calculated for each charged Higgs mass and value of $\tanb$.
The resulting \mbox{$5\,\sigma$} discovery contour is presented and discussed
in the next section.

\begin{figure}[]
    \begin{minipage}[t]{1\textwidth}
      \begin{minipage}[t]{.45\textwidth}
        \begin{center}
          \epsfig{file=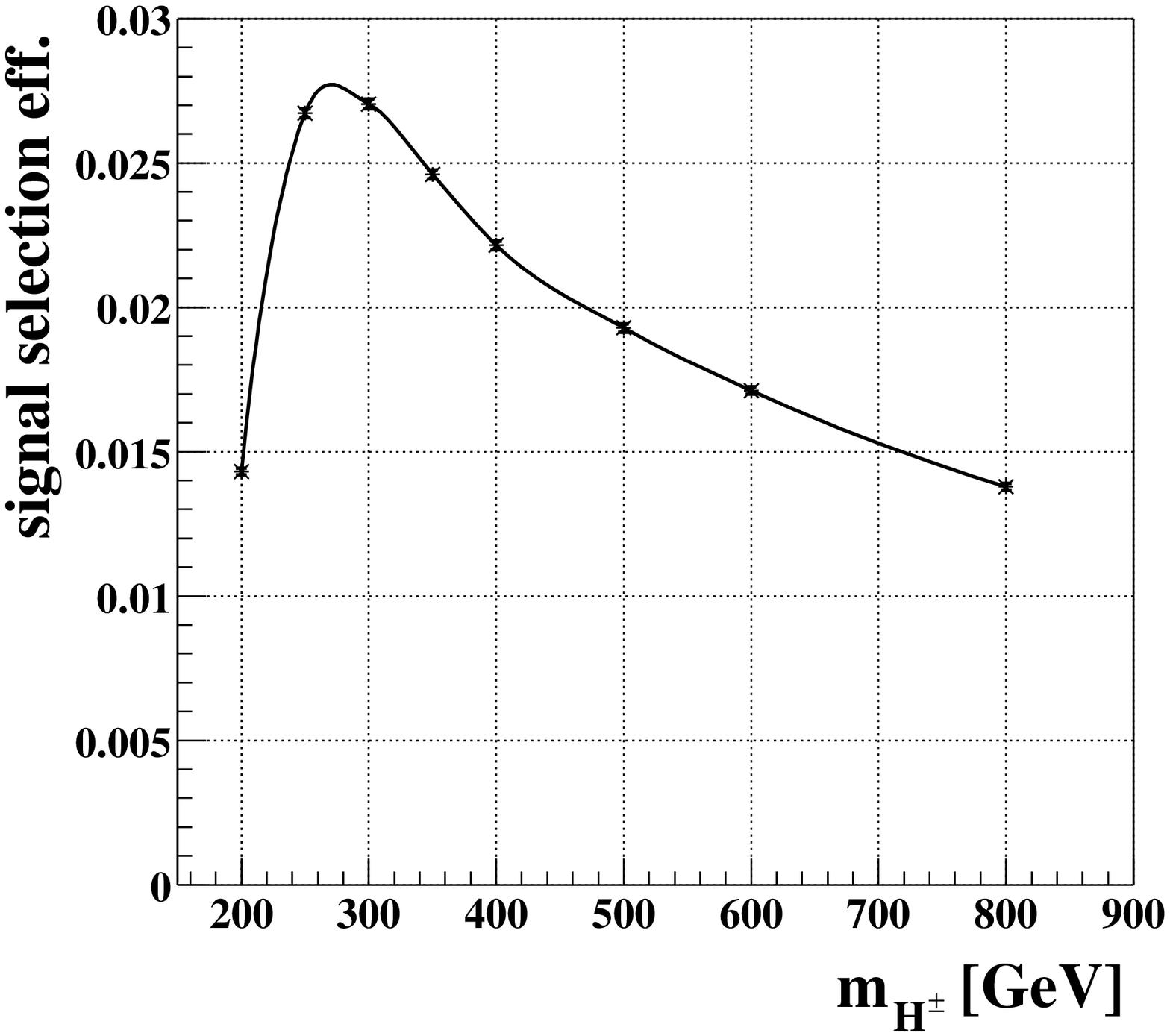,width=1.06\textwidth}
        \end{center}
      \end{minipage}
      \begin{minipage}[t]{.45\textwidth}
        \begin{center}
          \epsfig{file=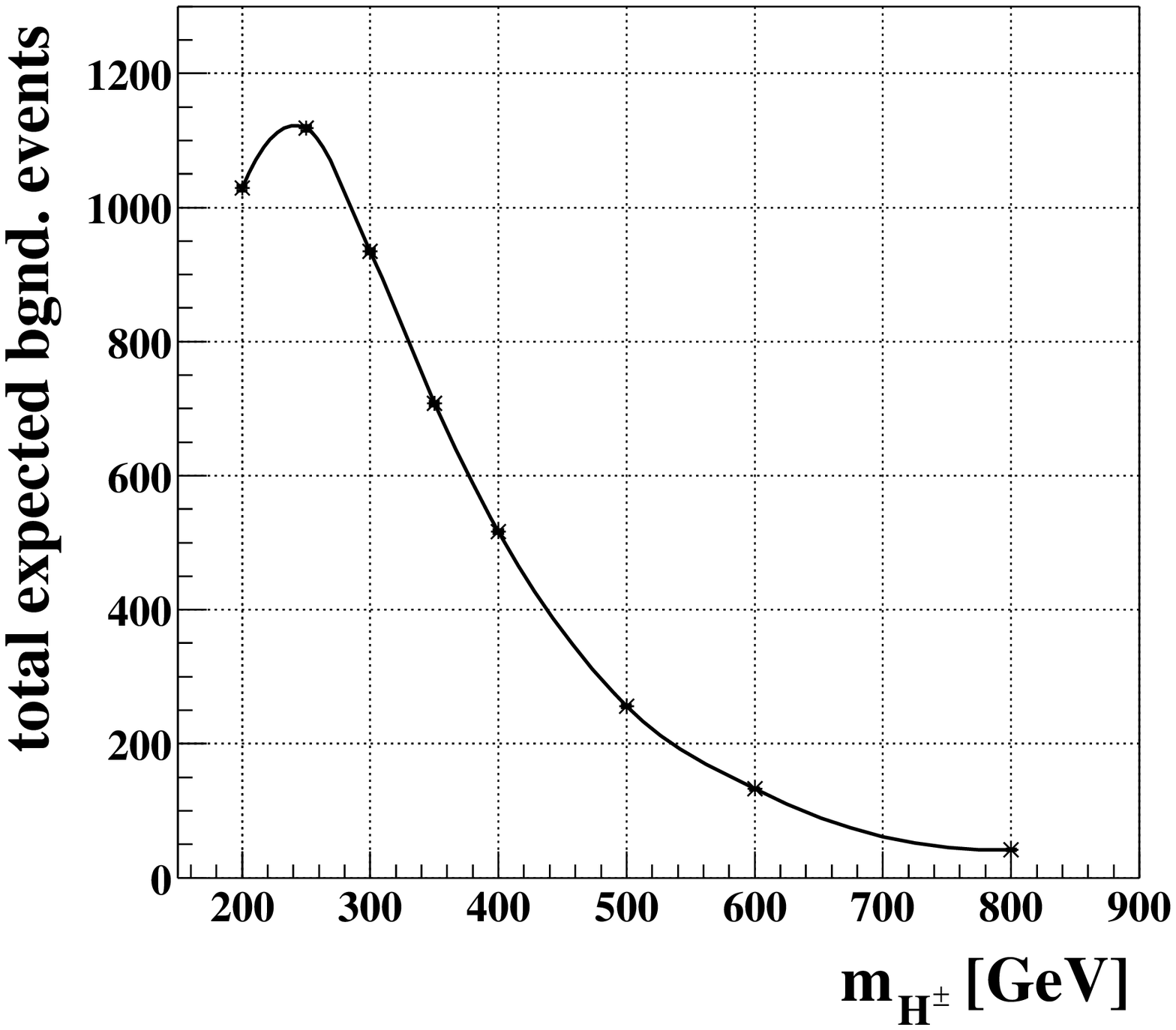,width=1.06\textwidth}
        \end{center}
      \end{minipage}
    \end{minipage}
\caption[]{\label{fig:selEff}\sl 
The signal selection efficiency and the total expected
number of background events after all cuts in the mass window of 
\mbox{$\pm 100\,\G$} 
assuming an integrated luminosity of \mbox{$30\,\ifb$}.
The dots represent the charged Higgs masses that were studied, 
and the line is a smooth interpolation.
}  
\end{figure}

Figure~\ref{fig:selEff} summarises the 
signal selection efficiency and the expected number of background
events after all cuts have been made for the whole range 
of charged Higgs masses studied.
The signal selection efficiency lies within \mbox{$1.4 - 2.7\,\%$} and
reaches its maximum around \mbox{$\mHpm\approx 300\,\G$}.


\section{Results} 
\label{sec:results}
\begin{figure}[]
  \begin{center}
    \epsfig{file=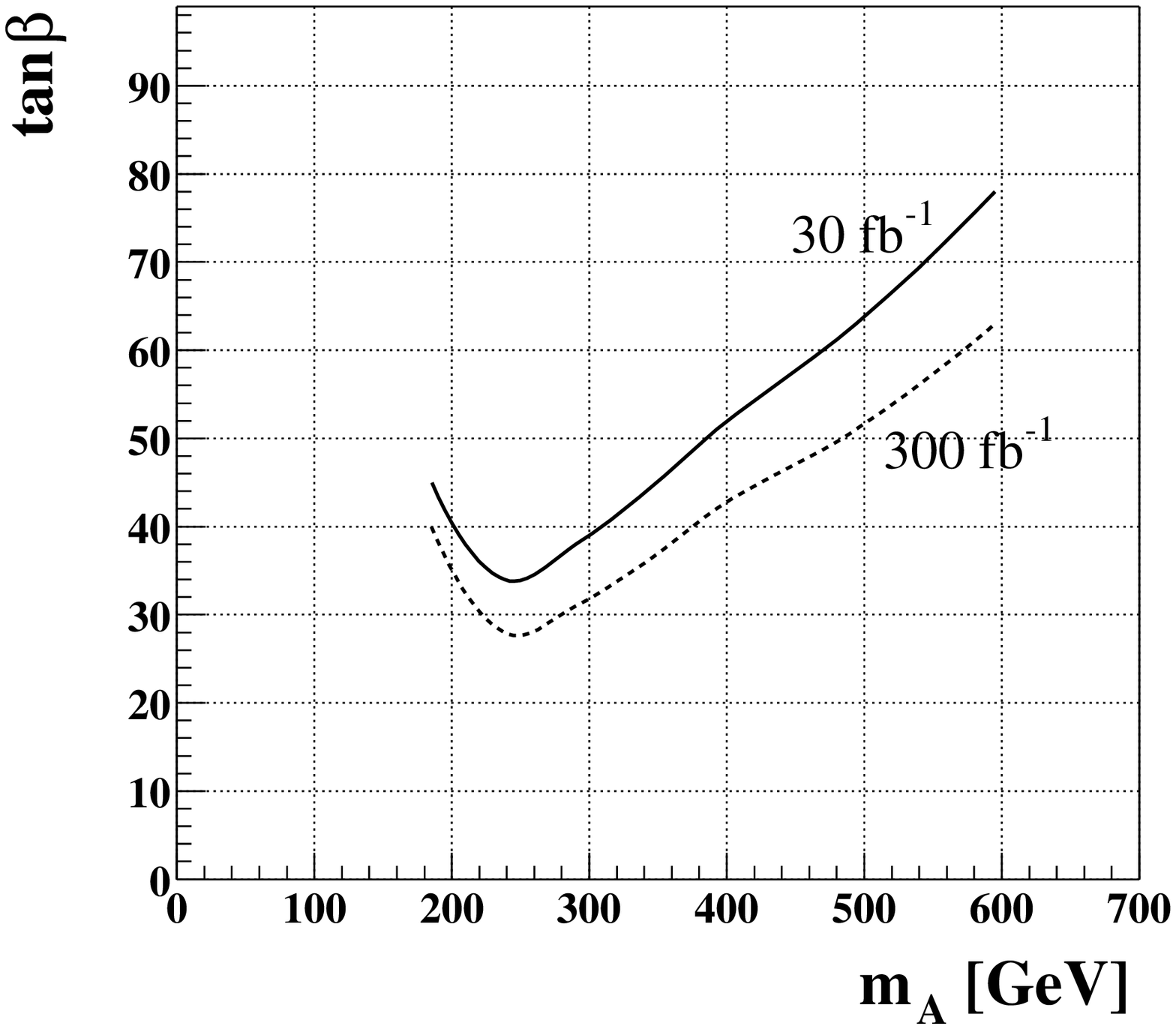,width=.8\textwidth}
  \end{center}
\caption[]{\label{fig:discPlaneNOSYST}\sl 
The \mbox{$5\,\sigma$} discovery contours for the process
\mbox{$gg\rightarrow tb\Hpm\rightarrow	
t\bar{t}b\bar{b}\rightarrow	
b\bar{b}b\bar{b}l\nu\bar{q}q'$} 
in the
\mbox{$(m_{\mathrm{A}},\tanb)$} plane for the low and the high luminosity
option of the \lhc, assuming integrated luminosities of \mbox{$30\,\ifb$} and
\mbox{$300\,\ifb$} respectively. A common renormalisation and factorisation
scale of \mbox{$\mu_F^2=\mu_R^2=(m_{T}^2(t)+m_T^2(b))/2$}
and a running $b$--quark
mass are assumed when evaluating the cross sections of the various
processes involved. No systematic uncertainties are taken into
account.
}  
\end{figure}

The results of the analysis are described in this section in terms of 
\mbox{$5\,\sigma$} discovery contours in the 
\mbox{$(m_{\mathrm{A}},\tanb)$} plane. 
They are presented for integrated luminosities of \mbox{$30\,\ifb$} for the low
luminosity option and \mbox{$300\,\ifb$} for the high luminosity option of
the \lhc. 
In the latter case a $b$--tag efficiency of $\epsilon_b=0.5$ is assumed and the 
\mbox{$p_T^{\mathrm{min}}$} cut on all jets is raised to 
\mbox{$\pT^{\mathrm{jet}}>30\,\G$}.
The degradation of jet--energy measurements due to
pile--up is taken approximately into account by choosing the 
high luminosity option in the \atlfast\ simulation package. 
Figure~\ref{fig:discPlaneNOSYST} shows the expected discovery contours
taking no systematic uncertainties into account. 
The charged Higgs boson can be detected for \tanb\ values down to 
$35$ for \mbox{$\mHpm\approx 250\,\G$} based on an integrated luminosity
of \mbox{$30\,\ifb$}. For the high luminosity option and 
\mbox{$300\,\ifb$} the 
reach in \tanb\ goes down to approximately $28$ for the same charged
Higgs mass region.
The analysis presented here depends heavily on the $b$--tagging performance 
and the reconstruction of jets with a relatively low $p_T$. This explains
why only a small improvement in the discovery potential is observed 
when switching to the high luminosity option.

\begin{figure}[]
  \begin{center}
    \epsfig{file=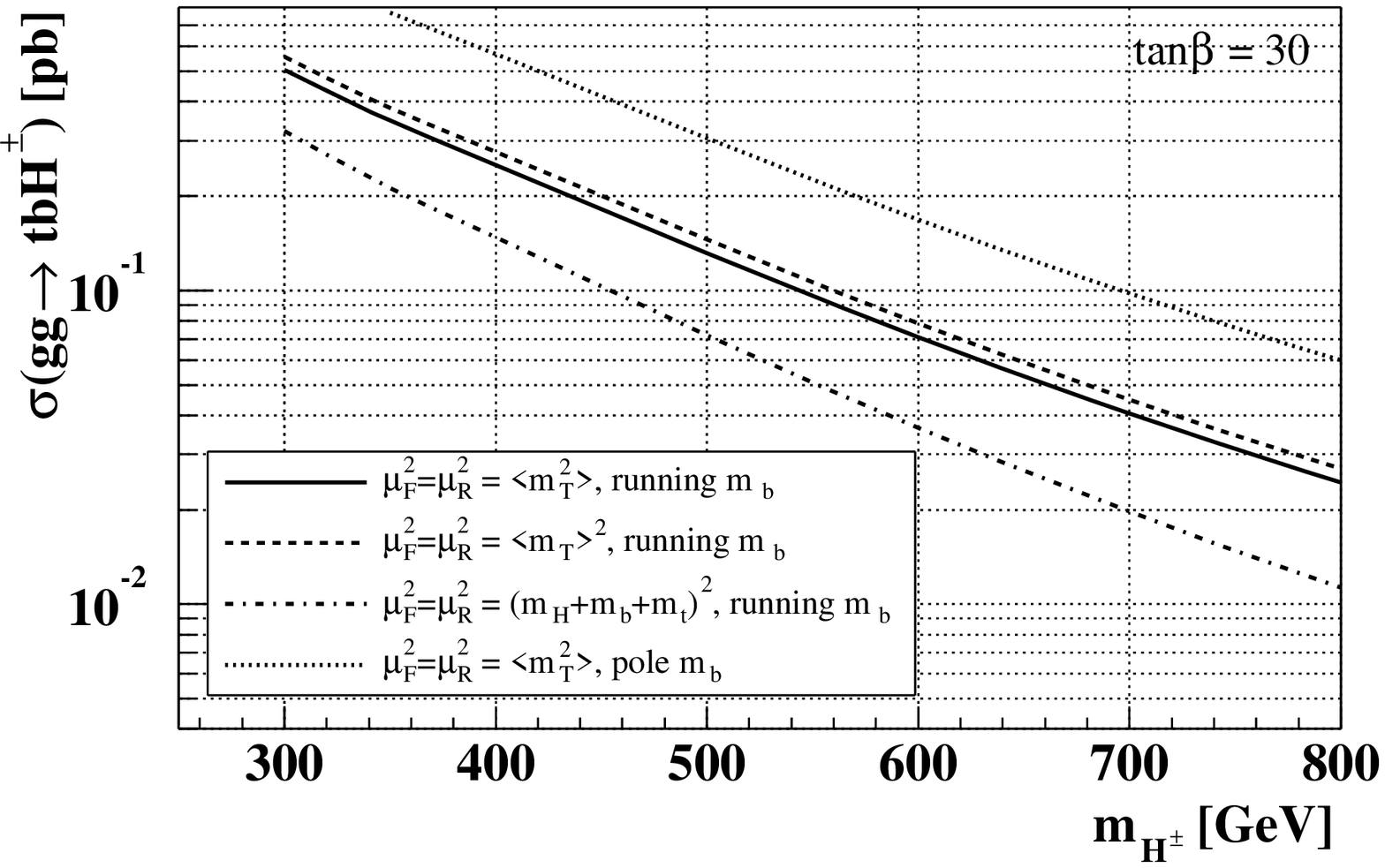,width=.8\textwidth}
  \end{center}
\caption[]{\label{fig:xsecsMarket}\sl 
cross section predictions for the \mbox{$gg\rightarrow tb\Hpm$} 
process obtained
with \herwig~$6.5$ assuming different QCD scales and $b$--quark mass 
evaluations. A value of $\tanb=30$ is chosen.
}  
\end{figure}

\begin{figure}[]
    \begin{minipage}[t]{1\textwidth}
      \begin{minipage}[t]{.45\textwidth}
        \begin{center}
          \epsfig{file=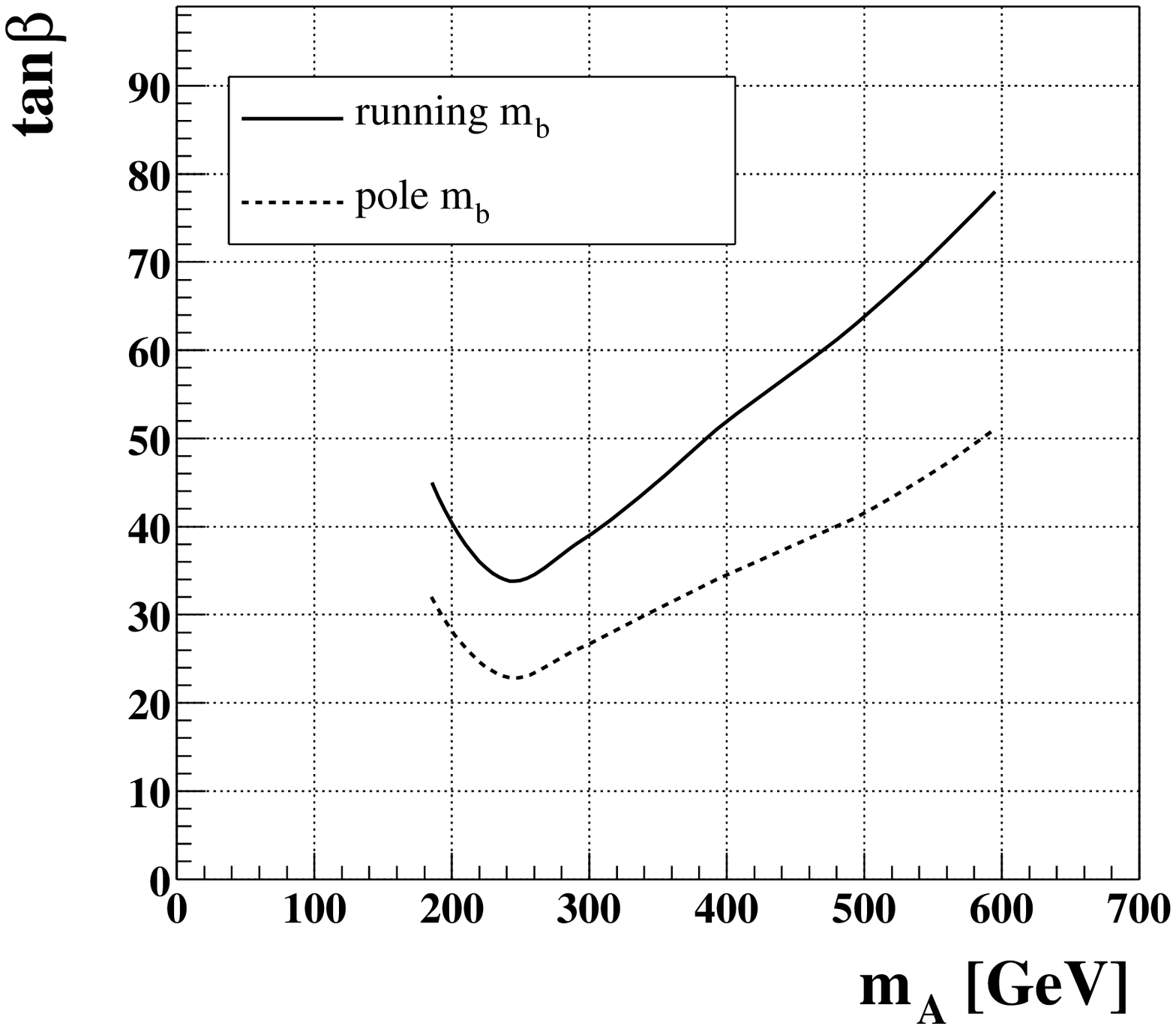,width=1.06\textwidth}
        \end{center}
      \end{minipage}
      \begin{minipage}[t]{.45\textwidth}
        \begin{center}
          \epsfig{file=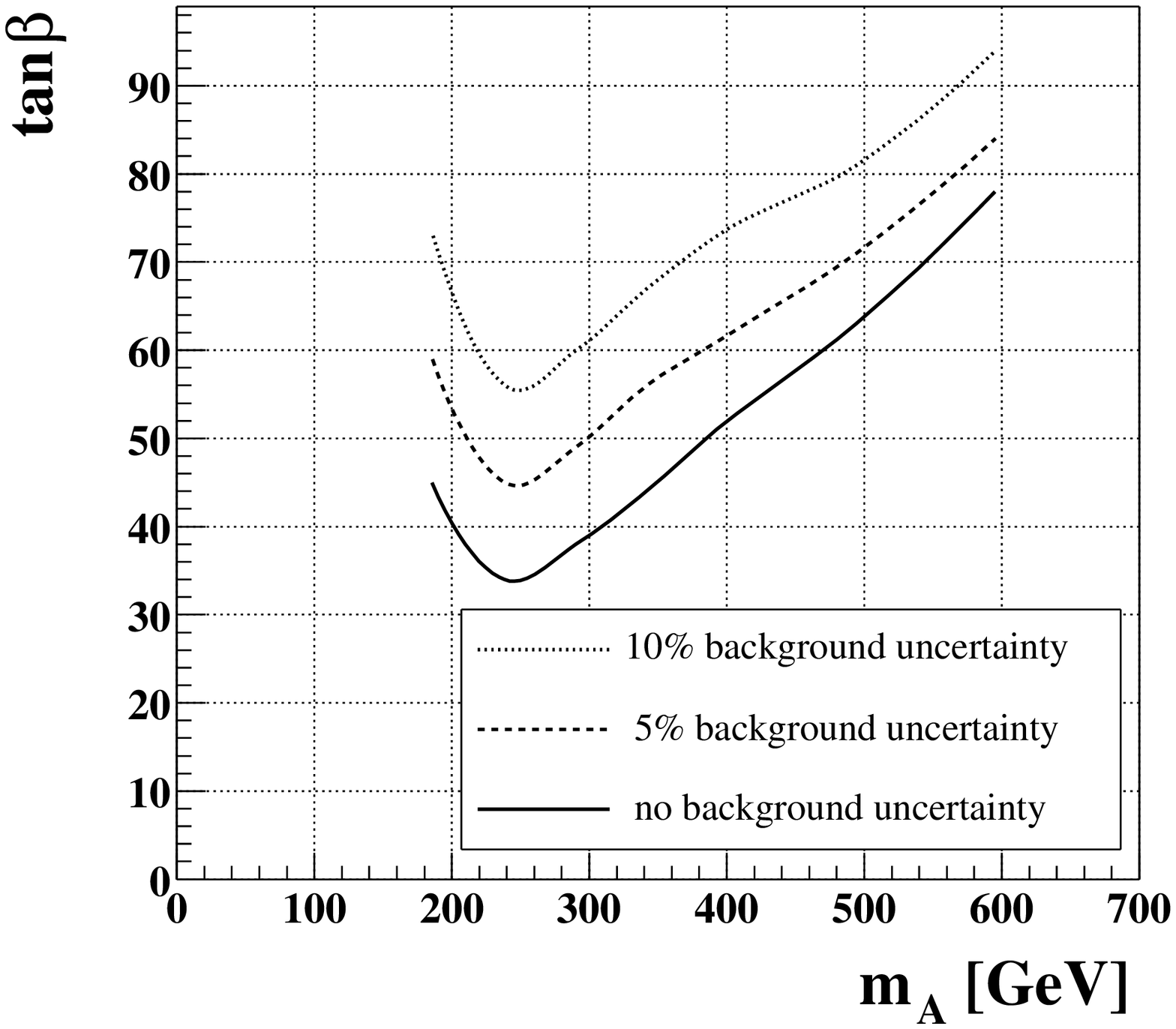,width=1.06\textwidth}
        \end{center}
      \end{minipage}
    \end{minipage}
\caption[]{\label{fig:discPlanePOLE}\sl 
Influences of systematic uncertainties on the discovery
potential assuming an integrated luminosity of \mbox{$30\,\ifb$} and 
\mbox{$\mu_F^2=\mu_R^2=(m_{T}^2(t)+m_T^2(b))/2$}. 
The left plot shows the improvement 
in discovery potential if a pole $b$--quark mass is assumed instead of a
running $m_b$ when evaluating
the \mbox{$gg\rightarrow tb\Hpm$} cross section.
The right plot shows \mbox{$5\,\sigma$} discovery contours taking systematic
uncertainties on the background cross section normalisation of $5\,\%$ and $10\,\%$
into account. Here a running $m_b$ is used.
}  
\end{figure}

The uncertainty in the prediction of the signal and background
cross sections due to the choice of QCD scale and 
running or pole $b$--quark masses is quite large 
as already discussed in section~\ref{sec:simulation}. 
Figure~\ref{fig:xsecsMarket} shows some cross section predictions
obtained with \herwig~6.5 for the \mbox{$gg\rightarrow tb\Hpm$} process,
assuming different QCD scales and $m_b$ evaluations.
The lower three curves represent cross section predictions for a running
$b$--quark mass and renormalisation and factorisation scales of
\mbox{$\mu_F^2=\mu_R^2
=(m_{T}(t)+m_T(b))^2/4$},
\mbox{$(m_{T}^2(t)+m_T^2(b))/2$}, and
\mbox{$(\mHpm+m_t+m_b)^2$}.
Whereas only a small difference is observed between the predictions of the
first two choices,
a rather large reduction in the expected signal cross section is
observed if a QCD scale of \mbox{$\mHpm+m_{b}+m_{t}$} is assumed. 
However, NLO calculations for the \mbox{$2\rightarrow 2$} process 
\mbox{$gb\rightarrow t\Hpm$} \cite{Plehn:2002vy} and the 
\mbox{$gg\rightarrow t\bar{t}H$} process \cite{gg_ttH} show that
this choice of scale might be too high. 
The same studies show also that the cross sections are likely to be
overestimated when using a pole $b$--quark mass. We therefore adopt a
running $b$--mass, ensuring K--factors larger than~1.
To illustrate the effect of a larger signal cross section prediction
we nevertheless 
show the cross sections expected 
when assuming a $b$--quark pole mass in the uppermost curve 
in Figure~\ref{fig:xsecsMarket} and the 
corresponding improvement in the discovery contour in 
Figure~\ref{fig:discPlanePOLE} (left plot).
The latter plot shows that
improvements in the
discovery potential due to
K--factors $>1$ might be sizable. 

The main \mbox{$gg / qq\rightarrow t\bar{t}b\bar{b}$} background
cross section prediction is also very sensitive to the
QCD scale~\cite{Kersevan:2002dd} and 
the uncertainties on the cross section prediction 
are of the same order as for the signal process. 
However, here we will assume that the background cross section can be
measured using side--bands in the reconstructed mass distribution
which are relatively signal--free. 
The precision of this procedure depends on the charged Higgs mass and
on the integrated luminosity available. 
No detailed study is conducted here, but to give some 
indication of how the discovery potential is affected by this
uncertainty on the expected \SM\ background, 
we assume an uncertainty of $5$--$10\,\%$ in the background normalisation, 
guided by the studies done
in~\cite{jochen}.
If the background MC samples produced with \acermc\ are passed to 
\pythia\ for string fragmentation and hadronisation instead of
{\herwig}'s cluster fragmentation, differences between \mbox{$5\,\%$} and
\mbox{$10\,\%$} are observed in the background prediction, depending on the
charged Higgs mass.

To illustrate the effects of uncertainties on the \SM\ background
prediction of this order of magnitude, we show the effect of
\mbox{$5\,\%$} and \mbox{$10\,\%$}
 uncertainties on the discovery potential in
Figure~\ref{fig:discPlanePOLE}.
Again, the corrections are found to be sizable.

\section{Conclusion and Outlook}
\label{sec:conclusion}
This note analyses the discovery potential for a charged Higgs boson
heavier than the top quark produced in the \mbox{$2\rightarrow 3$} process  
\mbox{$gg\rightarrow tb\Hpm$}.
The subsequent decay to heavy quarks \mbox{$\Hp\rightarrow t\bar{b}$} is
considered, leading to a final state consisting of four $b$--jets, two
light jets and one electron or muon plus missing energy. 
The whole production and decay chain reads 
\mbox{$gg\rightarrow tb\Hpm\rightarrow	
t\bar{t}b\bar{b}\rightarrow	
b\bar{b}b\bar{b}l\nu\bar{q}q'$}.
Studying the \mbox{$2\rightarrow 3$} process offers the possibility to detect
four $b$--jets in the final state and thereby reduce the \SM\
background considerably compared to case in which the \mbox{$2\rightarrow 2$}
production process \mbox{$gb\rightarrow t\Hpm$} is considered.

One of the main difficulties to overcome when trying to reconstruct
signal events is the high number of possible combinations of paired
reconstructed objects in order to reconstruct the charged Higgs boson. 
It is shown that the reconstruction is possible by employing multivariate
techniques, in which angular correlations are also taken into account.
By employing a likelihood selection separating the \SM\ background
from signal events, it is possible to detect the charged Higgs boson
in the 
\mbox{$gg\rightarrow tb\Hpm\rightarrow 
t\bar{t}b\bar{b}\rightarrow
b\bar{b}b\bar{b}l\nu\bar{q}q'$} channel 
down to values of \mbox{$\tanb\approx 28$} for charged Higgs masses around
\mbox{$250\,\G$} assuming an integrated luminosity of 
\mbox{$300\,\ifb$}. 
However, the theoretical uncertainties related to the cross section
predictions for both the signal process and the main \SM\ background
are quite large and lead to a sizable uncertainty in the expected
discovery contour in the \mbox{$(m_{\mathrm{A}},\tanb)$} plane.
NLO corrections to the signal cross section might result in an
improvement in the discovery potential whereas expected uncertainties
when measuring the \SM\ background contribution will degrade the
result.
We present the \mbox{$5\,\sigma$} discovery contour in
Figure~\ref{fig:discPlaneNOSYST} using a running $b$--quark mass and 
\mbox{$\mu_F^2=\mu_R^2=(m_{T}^2(t)+m_T^2(b))/2$} and taking no systematic
uncertainties on the background normalisation into account.

The goal of this analysis was to utilise the detection of the
fourth $b$--jet in the signal process in order to extend the discovery 
region
for the charged Higgs boson at high charged Higgs masses as suggested
in~\cite{Miller:1999bm}. 
This analysis shows that the encouraging 
results obtained in~\cite{Miller:1999bm}
do not hold when detector effects and mis--tagging of $b$--jets
are more properly taken into account. 

A direct comparison to a previous analysis~\cite{Assamagan:2000uc}
where the \mbox{$2\rightarrow 2$}
production process \mbox{$gb\rightarrow t\Hpm$} was used to produce a heavy
charged Higgs boson which subsequently also decays to heavy quarks, 
\mbox{$\Hp\rightarrow t\bar{b}$}, 
is not
possible at this stage, 
since the cross section predictions and production mechanisms 
for the \SM\ backgrounds that are assumed in the two cases are different.

Finally it should be noted that the results presented here might be
subject to another large systematic effect. As was mentioned in 
section~\ref{sec:simulation}, 
the $b$--tag efficiencies and rejection factors assumed are static,
i.e.\@ they do not depend on $\eta$ nor on the $p_T$ of the jet 
under consideration. 
This is clearly a rather crude approximation especially 
in the present analysis for which the detection of $4$ $b$--jets is
crucial.  More reliable results should be possible in the future using
a more accurate ($p_T$,$\eta$)--dependent parametrisation for the
$b$--tagging efficiency.

\begin{acknowledgments}
This work has been performed within the \atlas\ Collaboration, 
and we thank collaboration members for helpful discussions.
We have made use of the physics analysis framework and tools which 
are the result of collaboration--wide efforts.
Further we would like to thank Stefano Moretti, Johan Rathsman and
Gunnar Ingelman for helpful advice and discussions.  Parts of the MC
samples were produced on the Nordugrid~\cite{nordugrid} and we thank 
the development team for their support.
\end{acknowledgments}

\bibliographystyle{utphys}
\bibliography{gg_to_tbH}
\cleardoublepage
\end{document}